\documentclass[a4paper,fleqn]{cas-sc}

\usepackage[numbers]{natbib}

\usepackage{amsmath}
\usepackage{hyperref}
\usepackage{graphicx}			
\usepackage{microtype}
\usepackage{color}
\usepackage{xcolor}
\usepackage{float}
\usepackage{tikz}
\usepackage{mathtools}
\usepackage{bm}

\usepackage{soul}

\usepackage{algorithmic}
\usepackage{longtable}
\usepackage{colortbl}
\usepackage{multirow}
\usepackage[section]{placeins}
\usepackage{booktabs}

\graphicspath{{figures/},.}
\newcommand{\ie}{i.e.,}
\newcommand{\eg}{e.g.,}

\def\tsc#1{\csdef{#1}{\textsc{\lowercase{#1}}\xspace}}
\tsc{WGM}
\tsc{QE}
\tsc{EP}
\tsc{PMS}
\tsc{BEC}
\tsc{DE}

\begin{document}
\let\WriteBookmarks\relax
\def\floatpagepagefraction{1}
\def\textpagefraction{.001}


\shorttitle{Explaining the distribution of energy consumption at slow charging infrastructure for electric vehicles from socio-economic data}
\shortauthors{Straka, M. et~al.}

\title [mode = title]{Explaining the distribution of energy consumption at slow charging infrastructure for electric vehicles from socio-economic data}

\tnotemark[1,2]
\tnotetext[1]{This document is a result of the research project VEGA 1/0089/19 ``Data analysis methods and decisions support tools for service systems supporting electric vehicles''. It was partly supported by VEGA 1/342/18 ``Optimal dimensioning of service systems'', APVV-19-0441 ``Allocation of limited resources to public service systems with conflicting quality criteria'', Operational Program Research and Innovation in the frame of the project: ICT products for intelligent systems communication, code ITMS2014+ 313011T413, co-financed by the European Regional Development Fund and by the Slovak Research and Development Agency under the contract no. SK-IL-RD-18-005.}

\author[1,2]{Milan Straka}[type=editor,
auid=000,bioid=1,
role=Researcher,
orcid=0000-0002-8558-0408]
\fnmark[1]
\ead{milan.straka92@gmail.com}
\credit{Data curation, Methodology, Formal analysis, Investigation, Software, Validation, Visualization, Writing -- original draft}
\address[1]{Department of Mathematical Methods and Operations Research,
	University of \v{Z}ilina, Univerzitn\'{a} 8215/1, \v{Z}ilina, Slovakia}
\address[2]{University Science Park,
	University of \v{Z}ilina, Univerzitn\'{a} 8215/1, \v{Z}ilina, Slovakia}

\author[3,4,5]{Rui Carvalho}[type=editor,
auid=000,bioid=1,
role=Co-ordinator,
orcid=0000-0002-3279-4218]
\credit{Conceptualization, Methodology, Investigation, Resources, Supervision}
\address[3]{Department of Engineering, Durham University, Lower Mountjoy, South Road, Durham, DH1~3LE, UK}
\address[4]{Durham Energy Institute, Durham University, South Road, Durham, DH1~3LE, UK}
\address[5]{Institute for Data Science, Durham University, South Road, Durham, DH1~3LE, UK}

\author[6]{Gijs van der Poel}[type=editor,
auid=000,bioid=1,
role=Co-ordinator,
orcid=0000-0001-5525-9974]
\credit{Conceptualization, Resources}
\address[6]{ElaadNL, Utrechtseweg 310 (bld. 42B), 6812 AR Arnhem (GL), The Netherlands}

\author
[1,7]{{\v{L}}ubo\v{s} Buzna}[type=editor,
auid=000,bioid=1,
role=Co-ordinator,
orcid=0000-0002-5410-6762]
\credit{Conceptualization, Methodology, Investigation, Software, Project administration, Resources, Supervision, Visualization, Writing -- original draft, Writing -- review \& editing}
\address[7]{Department of International Research Projects - ERAdiate+,
University of \v{Z}ilina, Univerzitn\'{a} 8215/1, \v{Z}ilina, Slovakia}
\fntext[1]{Corresponding author.}

\begin{abstract}
Electric vehicles (EVs) belong to technologies alleviating concerns around energy security and greenhouse gas emissions. Recent developments and available forecasts suggest the continuous growth of the EV market share. However, the deployment of EVs on a large scale is associated with significant challenges. An adequate response to the forthcoming policy, technical, environmental, and planning issues implies the need for methods providing reliable decision support.

Here, we develop a data-centric approach enabling to analyse which activities, function, and characteristics of the environment surrounding the slow charging infrastructure impact the distribution of the electricity consumed at slow charging infrastructure. To gain a basic insight, we analysed the probabilistic distribution of energy consumption and its relation to indicators characterizing charging events. We collected geospatial datasets and utilizing statistical methods for data pre-processing, we prepared features modelling the spatial context in which the charging infrastructure operates. To enhance the statistical reliability of results, we applied the bootstrap method together with the Lasso method that combines regression with variable selection ability. We evaluate the statistical distributions of the selected regression coefficients. We identified the most influential features correlated with energy consumption, indicating that the spatial context of the charging infrastructure affects its utilization pattern. Many of these features are related to the economic prosperity of residents. Application of the methodology to a specific class of charging infrastructure enables the differentiation of selected features, e.g. by the used rollout strategy. 

Overall, the paper demonstrates the application of statistical methodologies to energy data and provides insights on factors potentially shaping the energy consumption that could be utilized when developing models to inform charging infrastructure deployment and planning of power grids.
\end{abstract}


%
%
%
%
%
%
%


\maketitle

\section{Introduction}
\label{sec:intro}
The European Union (EU) is moving towards commitments adopted under the Paris Agreement by aiming at domestic $CO_2$ cuts of at least $40$\% below $1990$ levels by $2030$~\cite{JRC_web}. To tackle this challenge, the deployment of plug-in hybrid electric vehicles (PHEVs), battery electric vehicles (BEVs) and fuel cell electric vehicles (FCEVs) appears to be inevitable~\cite[p.~91]{JRCreport_2019}. Electric mobility is growing at a rapid speed. In~$2018$, the number of new electric car sales almost doubled compared to~$2017$, and the global electric car fleet exceeded~$5.1$ million~\cite[p.~33]{globalEVOutlook_2019}. The world's largest electric car market is the People's Republic of China, followed by the EU and the United States (US). The global leaders in terms of electric car market share are Norway, Sweden, and the Netherlands, which is also dominant in terms of the charging infrastructure density, not only in Europe but also compared to the rest of the world~\cite[p.~4]{globalEVOutlook_2019}. In~$2018$, the global EV fleet consumed~$58$~TWh of electricity, which is comparable to the total electricity demand of Switzerland in $2017$~\cite[p.~9]{globalEVOutlook_2019}. The majority of outlooks envisions growing trend. In~$2030$, the global electric car sales are expected to reach~$23$ million, the stock will exceed~$130$ million vehicles (excluding two/three-wheelers), and electricity demand from EVs is estimated to reach almost~$640$~TWh~\cite[p.~6]{globalEVOutlook_2019}. In this scenario, slow chargers, which can provide flexibility services to power systems, are estimated to account for more than~$60$\% of the total electricity consumed globally to charge EVs.
\subsection{Motivation}
\label{subsec:motivation}
The recent developments in the deployment of electric vehicles are primarily affected by public policies implemented via economic stimuli and technological restrictions addressing industry, consumer behaviour and research activities~\cite{Bai17}. These efforts aim at influencing the course of events in the desired direction. As such, this is a complex process associated with a large number of decision-making situations and potentially leading to unexpected externalities. Hence, it is vital to have at hand tools which are able to address a broad range of decision problems linked to the deployment of EVs on a large scale.
\subsection{Literature review}
\label{subsec:previous_work}
%
%
%
%
Although electric vehicles have a longer history than fossil fuel vehicles, their mass adoption has started only recently~\cite{Chan_2013}. Initial barriers, such as range anxiety, low availability of charging infrastructure, high price and short lifetime of batteries, are slowly disappearing thanks to technological progress and policies promoting EVs. The impact of policies and the development of methods with the potential to facilitate the EV adoption has been a subject of intense research~\cite{Bai17,Xie17,Huang18,Sovacool18,Agugliaro17}. Specific methodologies have been developed to monitor and assess the behaviour of EV markets and EV components markets~\cite{Nilsson15,Obrecht19,LaClair19}. EV initiatives are largely motivated by sustainability issues. Hence, a reliable assessment of environmental benefits brought by EV adoption is of great importance~\cite{Hao17,Lave11}. Predicted implications of EV charging on the operation of power systems vary across the literature~\cite{Keoleian12,Sadeghianpourhamami_2018,Muratori18}. While some studies highlight positive impacts of reducing the peak power~\cite{Gonzalez18}, others report potentially negative effects, e.g. the level of detected load variability was concluded to be so high that it could potentially limit the integration of EV charging and renewable energy sources~\cite{Munkhammar18}. Underdeveloped charging infrastructure is another major factor hampering large scale EV adoption. For the past few years, the optimal placement and the sizing of EV charging stations have attracted significant attention of researchers~\cite{Dziurla14,Dou18,Zhao15}. The classification of available approaches is given in~\cite{Sanchari_2018}. Charging of EVs is not only influencing and influenced by the integration of renewable energy~\cite{Gerbaulet15,Qiu16}, but it enables bidirectional energy flows~\cite{Crawford18}, and it is expected to affect and be affected by energy markets~\cite{Samuelsen16} or by the way how EV drivers are navigated~\cite{Zhou19}.

%
%
With the growing number of EVs on roads in the last few years, a lot of research in the EV domain has been based on data-centric (or data science) methods and builds on available EV charging data.
A recent review~\cite{Pevec_2019} provides an overview of data sources and summarizes data science literature in the domain of EV charging. 

%
%
A few data-oriented studies have addressed the charging behaviour of EV drivers.
A set of key performance indicators characterizing utilization of chargers was defined and used to compare two rollout strategies: demand-driven and strategic rollout~\cite{Helmus_2018}. No rollout strategy is favourable over the other on all metrics, and the difference between strategies reduces as the EV adoption progresses. 
Using the location features, i.e. features characterizing the environment in which the infrastructure operates, authors in~\cite{Straka_2019} built prediction models for the popularity of charging infrastructure (i.e. the number of unique users). 
The multinomial logistic regression was used by~\cite{Wolbertus_2018} to determine key factors explaining heterogeneity in the charging duration of categorized charging sessions. The time-of-day-related variables and the type of charging station have the most substantial effect. Some location features such as type of the urban area, the density of chargers and parking possibilities were considered in this study as well. 
The ability of machine learning methods (random forest, gradient boosting and XGBoost) to predict the idle time (i.e. the time when an EV is connected to a charger without charging) was evaluated by~\cite{Lucas_2019}. The best model is XGBoost, reaching $R^2$ score of $60.3$\%. The most influential features are time-of-day-related features and the total energy supplied. Only one location feature, namely the type of the closest road segment, was considered. 

%
%
Analyses of charging infrastructure utilization focus on temporal aspects, typically applying time-series forecasting.
Predictability of energy consumption on single chargers was investigated in~\cite{Bikcora_2016}, finding potentially useful results only for some chargers.
Ref.~\cite{Louie_2017} identified and evaluated the time-series seasonal auto-regressive integrated moving average (ARIMA) models of EV load aggregated over $2400$ chargers. The long-term models (for two years) were found decidedly less accurate than the near-term models (for the most recent 60 weekdays and 24 weekend days).
Comparison of ARIMA models with decision trees, considering some exogenous features, concluded that former models are a better choice for forecasting aggregated EV charging loads~\cite{Buzna_2019}.
Considering four commonly used machine learning algorithms (K-nearest neighbour, pattern sequence-based forecasting, support vector regression and random forest), forecasts of EV charging load based on customer profile and charger measurements were compared by~\cite{Gadh16}, yielding similar prediction errors.
In~\cite{ai2018household}, the day-ahead EV charging load is forecasted as EV charging occurrence-time and the “no charge” day respectively, by several widely used machine learning algorithms. The best performance achieved the hybrid model combining Random Forest, Naive Bayes and XGBoost.

%
%
Spatial analyses of charging infrastructure utilization are less developed in the literature than temporal.
A preliminary exploratory analysis of spatial patterns formed by energy consumption on charging stations was presented in~\cite{Lucas_2018}. A study found a heterogeneous pattern, observing higher energy intensity in a small number of urban areas and $50$\% of the energy supplied comes from $19.6$\% of chargers.
Ref.~\cite{Pevec_2018} employs XGBoost model and five features describing the location and parameters of the charging infrastructure to predict its utilization. It also demonstrates how the model can be used to support decisions on locating the charging infrastructure at the level of zones with a radius of $3$~km.
\subsection{Our contribution}
\label{subsec:our_contribution}
In this paper, we develop a large scale spatial analysis of the energy consumption induced by charging of EVs. Our primary contribution is in extending the available knowledge of location factors, potentially affecting the energy consumption at public charging infrastructure. The candidate set of location factors is large. From the collected data, we extracted more than 120 features. We employ available statistical methods, specifically designed to explore a large set of potentially influential features. We focus on the slow charging infrastructure, which is believed to play a dominant role in the future~\cite[p.~6]{globalEVOutlook_2019}. As a case study, we consider the Netherlands. Recently, several prescriptive models appeared in the literature~\cite{Pevec_2018,Straka_2019,Wolbertus_2018,Lucas_2019} explaining some performance indicators of charging infrastructure from location features. For example, in~\cite{Pevec_2018} two such features are used, the number of points of interest and the number of competitive charging stations, to explain the utilization of charging infrastructure. In general, the decision which data to collect when developing a model is difficult. Our secondary contribution is in providing guidelines for these decisions when building prescriptive models involving energy consumption. Our last contribution is methodological. In the analysis, we consider the influence of multicollinearity and statistical stability of selected regression coefficients to improve the statistical reliability of results. In these ways, we contribute to methodologies for decision making support in domains of charging infrastructure deployment and power grids planning.
\section{Materials and Methods}
\label{sec:materials_and_methods}
\subsection{Terminology}
\label{subsec:terminology}
\begin{figure*}
	\centering
	\includegraphics[width=0.99\textwidth]{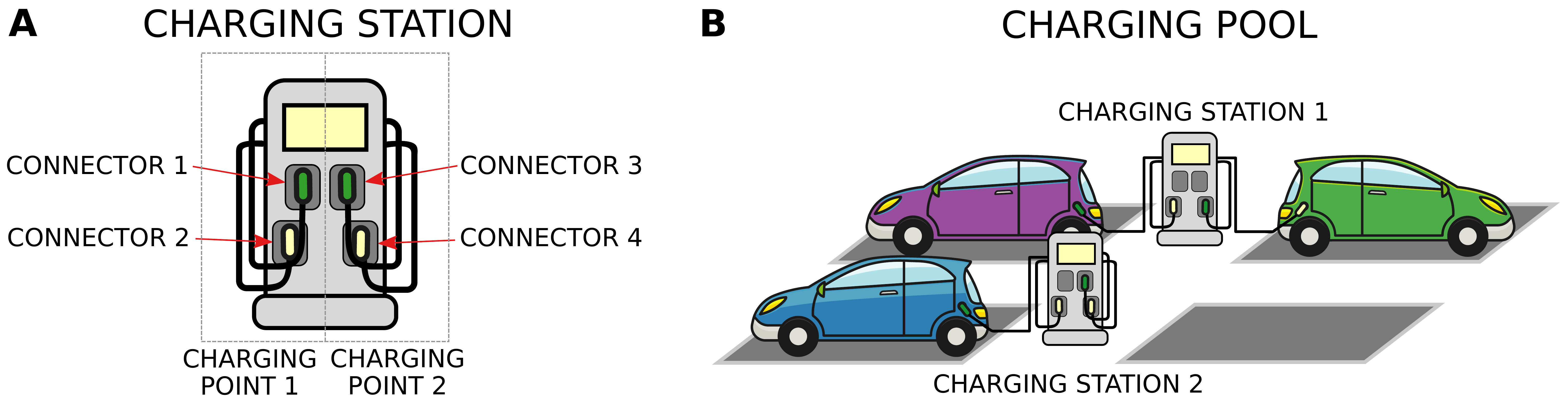}
	\caption{Schematics illustrating the terminology suggested in~Ref.~\cite{EVdefinitions} to denote the main components of the charging infrastructure for electric vehicles.}
	\label{fig:fig_terminology}
\end{figure*}
Charging of electric vehicles is a new field and various terminologies can be found in the literature. We follow Ref.~\cite{EVdefinitions}, where a \textit{connector} is defined as a physical interface between an electric vehicle and charging infrastructure through which electricity is delivered. Due to the incompatibilities of connector types used by different car manufacturers, several connectors might be available at a charging point, however, no more than one connector can be active at a time. A~\textit{charging point} is an energy delivery device equipped with one or more connectors. The charging point can charge an EV with a power which is less or equal to the maximum power, given in kW, referred to as \textit{the charging capacity}.
A~\textit{charging station} is composed of one or multiple charging points. A~user identification interface and all human-machine interfaces are attributed to the charging station and are shared by all charging points. A~\textit{charging pool} is one or a collection of charging stations including the adjacent parking lots. Components of the charging infrastructure are visualised in Figure~\ref{fig:fig_terminology}.
A~charging transaction starts by plugging a connector into an EV at a~\textit{start time} and ends by unplugging the EV at an \textit{end time}. The difference between end time and start time is \textit{the connection time}. A part of the connection time when EV was charging is referred to as \textit{the charging time}. \textit{The idle time} is the connection time minus the charging time.
\subsection{EVnetNL dataset}
\label{subsec:EvnetNL_dataset}
\begin{figure*}
	\centering
	\includegraphics[width=0.7\textwidth]{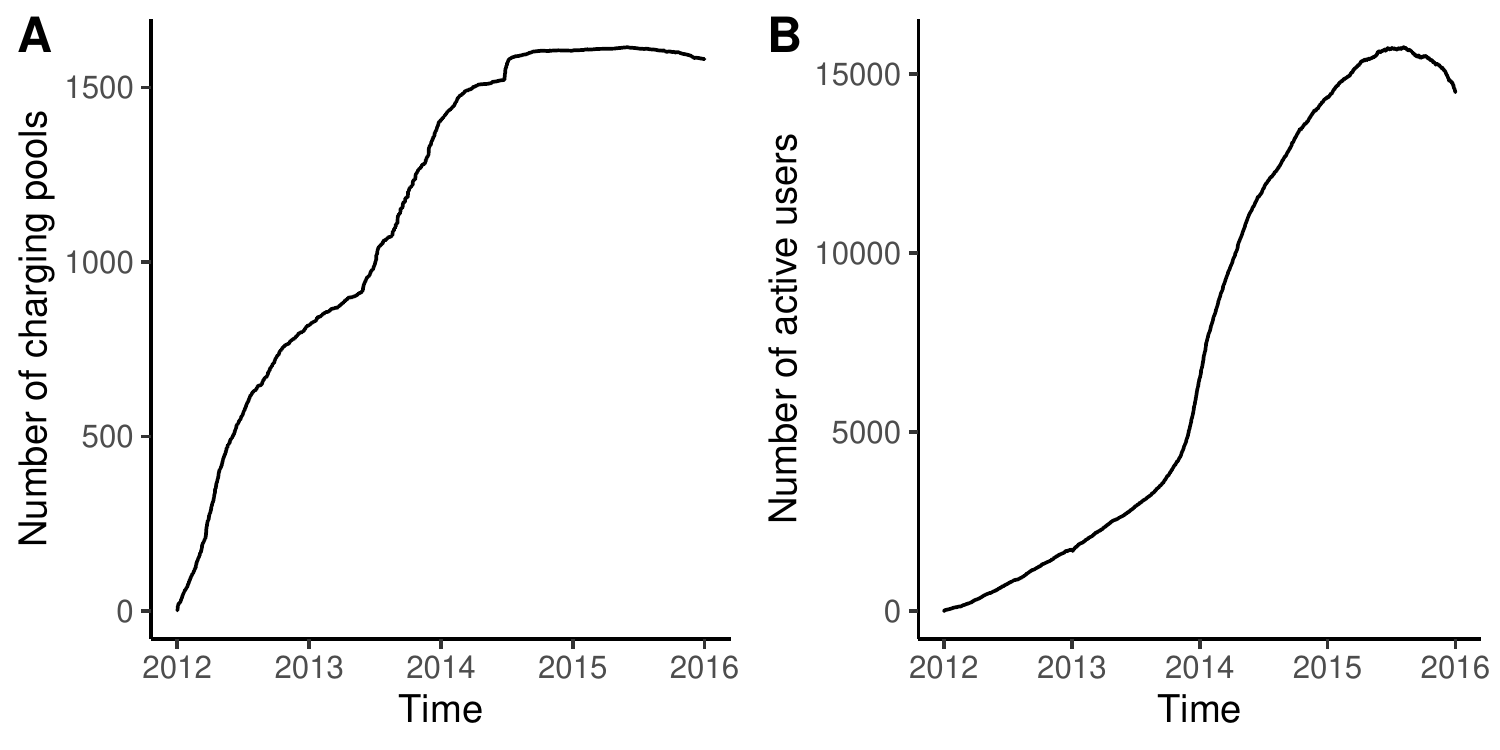}	
	\includegraphics[width=0.7\textwidth]{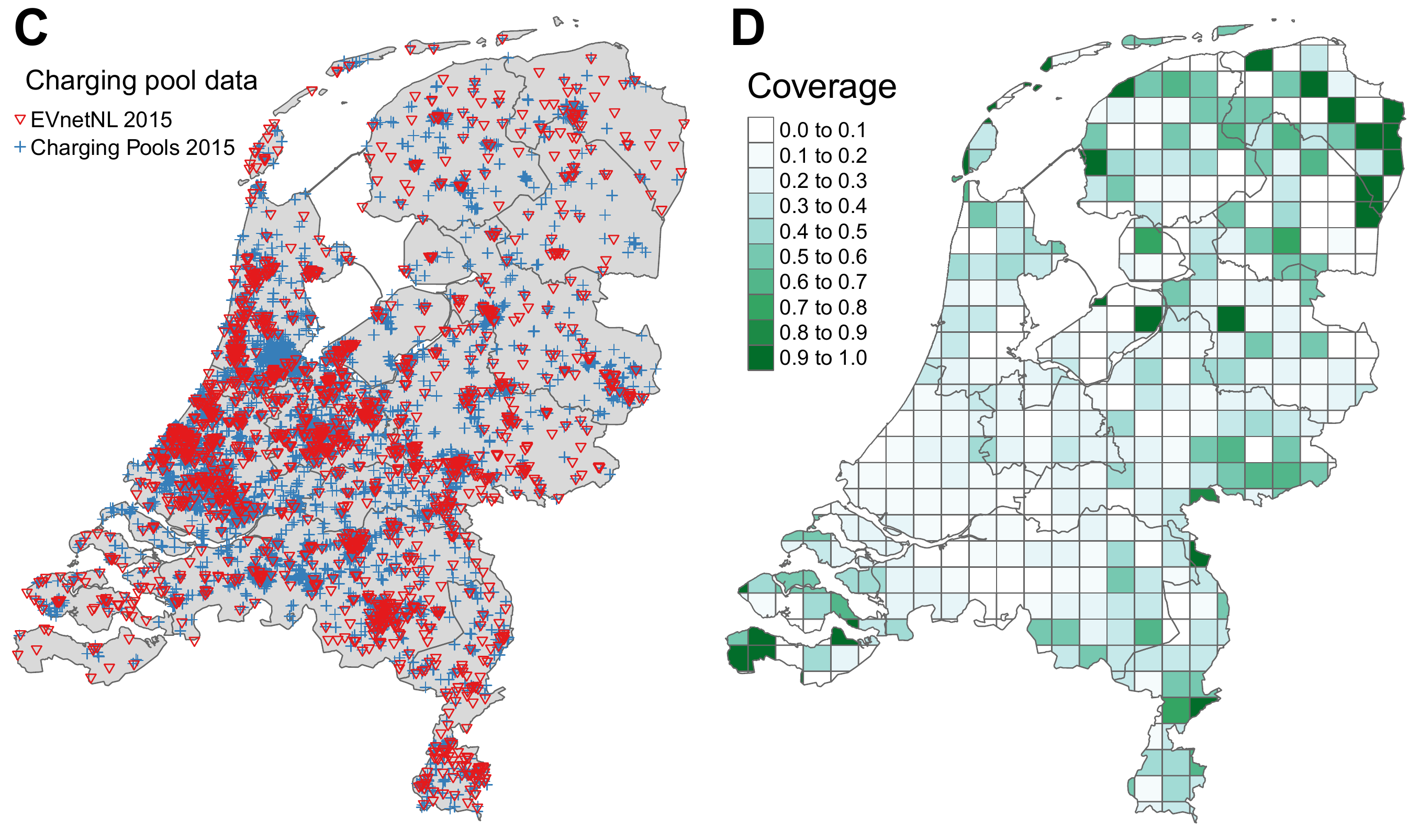}
	\caption{\textbf{A}~The number of charging pools in active use as a function of time. To estimate the number of active charging pools, the presence of each charging pool is bounded by the start time of the first charging transaction and the end time of the last charging transaction recorded in the EVnetNL dataset. \textbf{B}~The number of active users as a function of time (estimated from the first and the last recorded use of RFID cards). The number of active users is peaking in~$2015$ as in this period many RFID cards are used only a few times. \textbf{C}~A map of the Netherlands showing the geographical locations of EVnetNL charging pools operational in the year~$2015$ (triangles) together with the charging pools from the dataset Charging pools~$2015$ (crosses). In the Netherlands, $17~786$ slow charging points were operational in~$2015$, according to~\cite{eafo}. In the Charging pools~$2015$ dataset, we identified~$8~366$ unique positions of charging pools. Considering the distribution of charging points at charging pools observed in the EVnetNL dataset, we estimate that the Charging pools~$2015$ dataset covers about~$78.3$\% of all charging pools. \textbf{D}~The spatial representativeness of the EVnetNL dataset estimated by calculating the ratio between the number of stations in the EVnetNL and in the Charging pools~$2015$ datasets for square cells of a regular grid.}
	\label{fig:fig_2abcd}
\end{figure*}
	
\begin{figure*}
	\centering
	\includegraphics[width=0.7\textwidth]{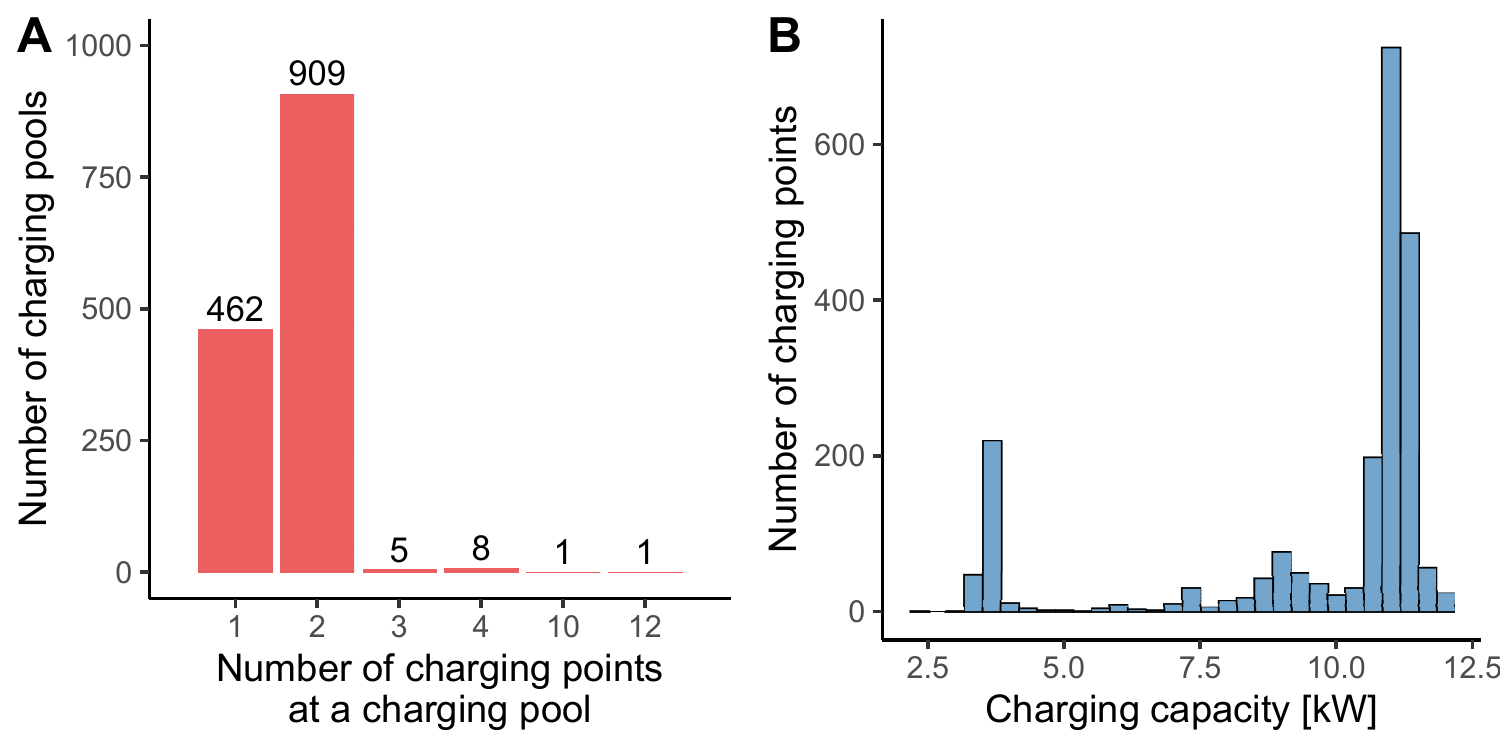}
	\caption{\textbf{A}~The number of charging pools in use on January~1,~2015 with a given number of charging points in the EVnetNL dataset. \textbf{B}~Histogram of charging capacities of charging points estimated from the meter reading values.}
	\label{fig:fig_1ab}
\end{figure*}
	
The EVnetNL dataset has been provided to us for research purposes by ElaadNL, a Dutch research organisation involved in the development, deployment and operation of EV charging infrastructure. The data is organised in two tables. The table \textit{Transactions} contains~$1~060~763$ rows, each characterising an individual charging event, with columns such as an identifier of a charging point, GPS coordinates of charging station, start time, end time, connection time, idle time, charging time, number of the used RFID cards, consumed energy and unique identifier of the charging event. The second table, \textit{Meterreadings}, has~$32~440~911$ rows, each corresponding to a meter reading that is taken each~$15$ minutes if a vehicle is connected. A meter reading is described by the identifier of the transaction, identifier of the charging point, UTC timestamp and value of the meter. 
	
The EVnetNL dataset covers~$1747$ charging stations equipped with~$2893$ charging points (identified by unique labels) that were operated by the ElaadNL in the period from January~2012 until March~2016. Some charging stations are located close to each other (e.g. in one parking lot). Consequently, the urban context of charging stations cannot be distinguished. Therefore, we considered the charging pools as the main object of the analysis. Locations of charging pools we estimated by aggregating the charging stations. For each charging station, we identified a set of neighbouring stations located within the radius of~$50$ meters. For every pair of stations distant more than~$50$ meters from each other, we found an empty intersection of sets of neighbouring stations. Therefore, we selected one representative charging station in each set of neighbouring stations, and all stations in the set were merged and formed a charging pool.

In the analysis of energy consumption, we consider only the period from January~1,~2015, until December~31,~2015, as this is the latest available complete year, when the number of charging pools was reasonably stable (see Figure~\ref{fig:fig_2abcd}A). As shown in Figure~\ref{fig:fig_2abcd}B, the number of active users, estimated from the number of RFID cards in use, exceeded the value~$15~000$ and it has been relatively stable throughout the year~2015 as well. We obtained a set of~$1604$ charging pools operational in~$2015$ while being distributed across the entire area of the Netherlands (see Figure~\ref{fig:fig_2abcd}C). In Figure~\ref{fig:fig_2abcd}D we analyse spatial representativeness of the EVnetNL dataset by calculating the ratio between the number of EVnetNL charging pools operational in~2015 and the number of charging pools in the Charging pools~2015 dataset (for more information about the dataset please refer to the Section~\ref{secSI:charging_pools_2015} of the SI file) for cells of a regular square grid. The ratio takes higher values in the east and south of the Netherlands. The EVnetNL dataset covers smaller cities better, while in large cities, such as Amsterdam and Rotterdam, only a small percentage of charging stations is covered.

We excluded~$40$ charging transactions for which either meter values or charging start and stop times were inconsistent across Meterreadings and Transactions tables. To eliminate charging pools with sparse usage patterns, we excluded from the analyses~$218$ pools with either less than~$30$ charging transactions in~2015 or less than~$1$ kW~charging capacity. To minimize the effects of the transition period that follows the introduction of a new charging pool, we consider only charging pools that have been in use before January~1,~2015 (we excluded~$87$ pools established in~2015 and later). After these rearrangements, we retained~$369~550$ transactions taking place on~$1~386$ EVnetNL charging pools. The large majority of charging pools possess~$1$ or~$2$ charging points and deliver power ranging up to~$12.5$~kW. Often, the fast charging is declared when the power is exceeding the value of~$22$~kW~\cite{EVdefinitions}, hence, all considered charging pools are used for slow charging (see Figure~\ref{fig:fig_1ab}A-B ).
\subsection{Geospatial datasets}
\label{subsec:databases}
\begin{figure*}
	\centering
	\includegraphics[width=0.97\textwidth]{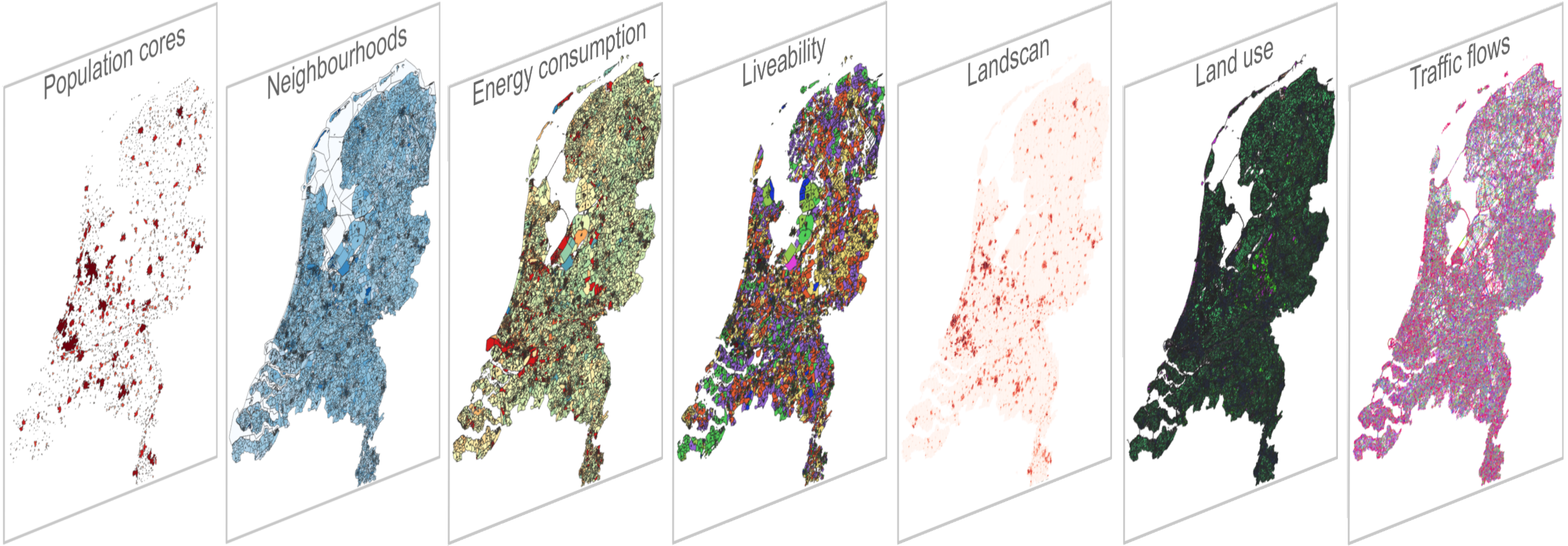}
	\caption{Overview of geospatial datasets that have been compiled to characterize the geographical area and human activities in the vicinity of charging pools.}
	\label{fig:fig_4}
\end{figure*}

To characterize the area and human activities taking place in the vicinity of charging pools distributed across the territory of the Netherlands, we collected potentially relevant publicly available geospatial datasets illustrated in Figure~\ref{fig:fig_4}. 

The geospatial datasets describe locations on the Earth's surface by geometric objects (points, polylines, polygons) and associate geometric objects with alphanumeric attributes. We inspected all available attributes and if multiple datasets contained the same or very similar data, we considered a more complete source or the source providing data with higher resolution. Moreover, we excluded attributes that lead to multicollinearity (e.g. from the triplet of attributes the total population, the population of men and the population of women, we considered only the first attribute as the populations of men and women tend to be very similar, hence, constitute one half of the total population). In what follows, we briefly describe each geospatial dataset. The complete list of selected attributes for each dataset is given in the SI file, Tables~\ref{tabSI:Population_cores_attr}-~\ref{tabSI:openchargemap}.
\subsubsection{Population cores}
The population cores are continuous spatial units with at least $25$~homes or $50$~registered residents~\cite{popcores}. The dataset associates the population cores with the information about the households (e.g. size and composition), the cardinality of population age groups, and family and civil status of residents. It also includes the information about the employment rate of residents, type of their occupation and characteristics of real-estate properties. The spatial resolution of this dataset is rather low (typically, a population core corresponds to a municipality) and we use it to investigate whether aggregate characteristics of municipalities can explain the usage pattern of charging pools. From this dataset, we selected~\textbf{45} attributes for further analysis (see Table~\ref{tabSI:Population_cores_attr} of the SI file).
\subsubsection{Neighbourhoods}
The neighbourhoods are spatial units that are approximately uniform when considering the type of built-up area or the socio-economic indicators~\cite{neighbpop}. The neighbourhoods are used by the Dutch national statistical office to collect, maintain and distribute the statistical data. In addition to the population data, the Neighbourhoods dataset maintains information about the number of address points within the neighbourhoods, level of the income of the residents, number of registered private and company vehicles and so on. From the available attributes, we selected \textbf{63} attributes listed in Table~\ref{tabSI:Neighbourhoods} of the SI file.
\subsubsection{Energy consumption}
The Energy consumption dataset contains records of the annual natural gas and electricity consumption in residential houses and industrial facilities, together with the number of buildings equipped with metering devices~\cite{energyatlas}. The spatial resolution is identical with the Neighbourhoods dataset. For further analyses, we selected \textbf{12} attributes that are detailed in Table~\ref{tabSI:energy_consumption} of the SI file.
\subsubsection{Liveability}
The Liveability dataset has been introduced by the Dutch Ministry of the Interior and Kingdom Relations to monitor the quality of living in Dutch neighbourhoods~\cite{liveability}. We use the liveability index 2016, that was revised in 2015. The liveability is quantified by a composite index and by five specific indices evaluating categories such as housing, socio-economic background of residents, services, safety, and the environment. Hence, from the liveability dataset, we extracted~\textbf{6} attributes listed in Table~\ref{tabSI:liveability} of the SI file.
\subsubsection{Landscan}
We use the LandScan~2015~\cite{landscan} high resolution population raster grid estimating the~24-hour average of population count with a spatial resolution of approximately $1~$km~$\times~1~$km. In contrast to the Population cores or the Neighbourhoods datasets that capture the residential population only, the Landscan considers the mobility of residents. 
\subsubsection{Land use}
The Land use dataset describes the occupation of land in the Netherlands by polygons~\cite{lccbs}. Each polygon is assigned an attribute value taking one out of predefined classes of land use. Examples of land use categories are traffic areas,building sites, recreational areas and business areas. Complete list of \textbf{25} categories is given in Table~\ref{tabSI:land_use} of the SI file.
\subsubsection{Traffic flows}
To model the impact of traffic on the usage of charging pools, we consider the traffic flows dataset~\cite{traffic_flows}. The dataset is organized around a high resolution model of the road network and the traffic flow information is added in the form of attributes that are associated with the road segments. The description of~\textbf{9} attributes that have been selected to compile features is given in Table~\ref{tabSI:traffic_flows} of the SI file.
\subsubsection{OpenStreetMap}
The OpenStreetMap (OSM) is one of the most successful free maps~\cite{osm}. From the OpenStreetMap of the Netherlands, we extracted all points of interest (POIs) considering $2$~km~$\times~2$~km squared areas centred at the positions of the EVnetNL charging pools. We identified $593$ different POI types, some of them appearing in only very few instances. For this reason, we associated manually POI types with one of the~\textbf{15} categories listed in Table~\ref{tabSI:openstreetmap} of the SI file. The POIs, organized in these $15$ categories, were used to model the venues in the proximity of charging pools that are often visited by EV drivers.
\subsubsection{Charging pools~2015}
Aiming at estimating the positions of all available charging pools present in the Netherlands by the end of the year~2015, the Charging pools 2015~dataset was compiled from the EVnetNL, OpenChargeMap~\cite{ocm} and OplaadPalen~\cite{opp} datasets. Utilizing the date when a charging station was added to the dataset, we extracted from the OpenChargeMap and OplaadPalen positions of all charging stations that were available by the end of~2015. As with the EVnetNL dataset, we estimated the position of charging pools from the positions of charging stations, while utilizing the information about the geographical proximity. In the first step, we added to the Charging pools~2015 all EVnetNL charging pools available in~2015. In the second step, we added one-by-one to the Charging pools~2015 dataset the charging stations from the OpenChargeMap and OplaadPalen datasets, if their position was more than~$50$ meters distant from already added pools. This way, we obtained positions of~$8~366$ charging pools (see Figure~\ref{fig:fig_2abcd}C).
\subsection{Data pre-processing}
\label{subsec:data_preprocessing}
To prepare the data for the analyses, we applied the pre-processing procedure composed of three stages: the missing value handling, the extraction of features and the analysis of potential data modelling problems, described in the following subsections.
 
\subsubsection{Missing values handling}
\label{sec_missing_value_handling}
Values of some attributes associated with geometric objects in geospatial datasets are missing. By visualising the geometric objects with missing attribute values on the map, we identified two main sources of problems. Some geometric objects with missing data represent water landscape. In such a case, we excluded the geometric objects from datasets. The second source of specific problems with missing data are cities of Baarle-Nassau and Baarle-Hertog with peculiar borders between the Netherlands and Belgium. In this case, we ignored missing values as the EVnetNL dataset does not feature any charging pool in this area. 
We applied a two-step approach to handle missing values (see section~\ref{subsecSI:handling} of the SI file). In the first step, when it was possible we estimated the missing values from the available data. Further analysis revealed that in the geospatial data the attribute values tend to be missing in areas with low intensity of human activities (e.g. low population, low density of buildings, low electricity consumption, etc.). Therefore, in the second step, we applied simple rules that set the missing values of some attributes to zero (or lowest possible value) in areas with low intensity of human activities. Remaining missing values are addressed after generating features characterising the vicinity of charging pools. 

\subsubsection{Preparation of features}
\label{preparation_of_features}
We modelled each charging pool as a single point defined by GPS coordinates. We used a buffer, circular area centred at the position of charging pool and having the radius~$r$, to model the vicinity of charging pools. Values of features were calculated from GIS polygon data, considering spatial intersections between the area of buffers and GIS geometric objects, while assuming a uniform spatial distribution of considered quantities over the area represented by GIS geometric objects. For POI data (OpenStreetMap and Charging Stations~2015), the features are defined two ways: as the distance from the pool to the closest point of interest and as the density of points of interest within the buffer area. We set the buffer radius to~$r~=~350$ meters. For more details please refer to Section~\ref{secSI:features} of the SI file.

To ensure that the handling of missing values cannot influence significantly the results of the analysis, we applied to each feature the following rule: The feature is used in the analysis if areas with missing values of the attribute take less than~$15$\% of the buffer areas, otherwise it is excluded. To make sure that features are not built based on GIS data with a large proportion of missing values, if there was less than~$1.5$\% of feature values missing after applying the estimations, missing values were imputed by a median value, otherwise, the feature was discarded. Finally, we obtained~$\boldmath{195}$ features. 

\subsubsection{Analysis of potential data modelling problems}
\label{treatment_of_data_problems}
\begin{figure}
	\centering
	\includegraphics[width=0.5\columnwidth]{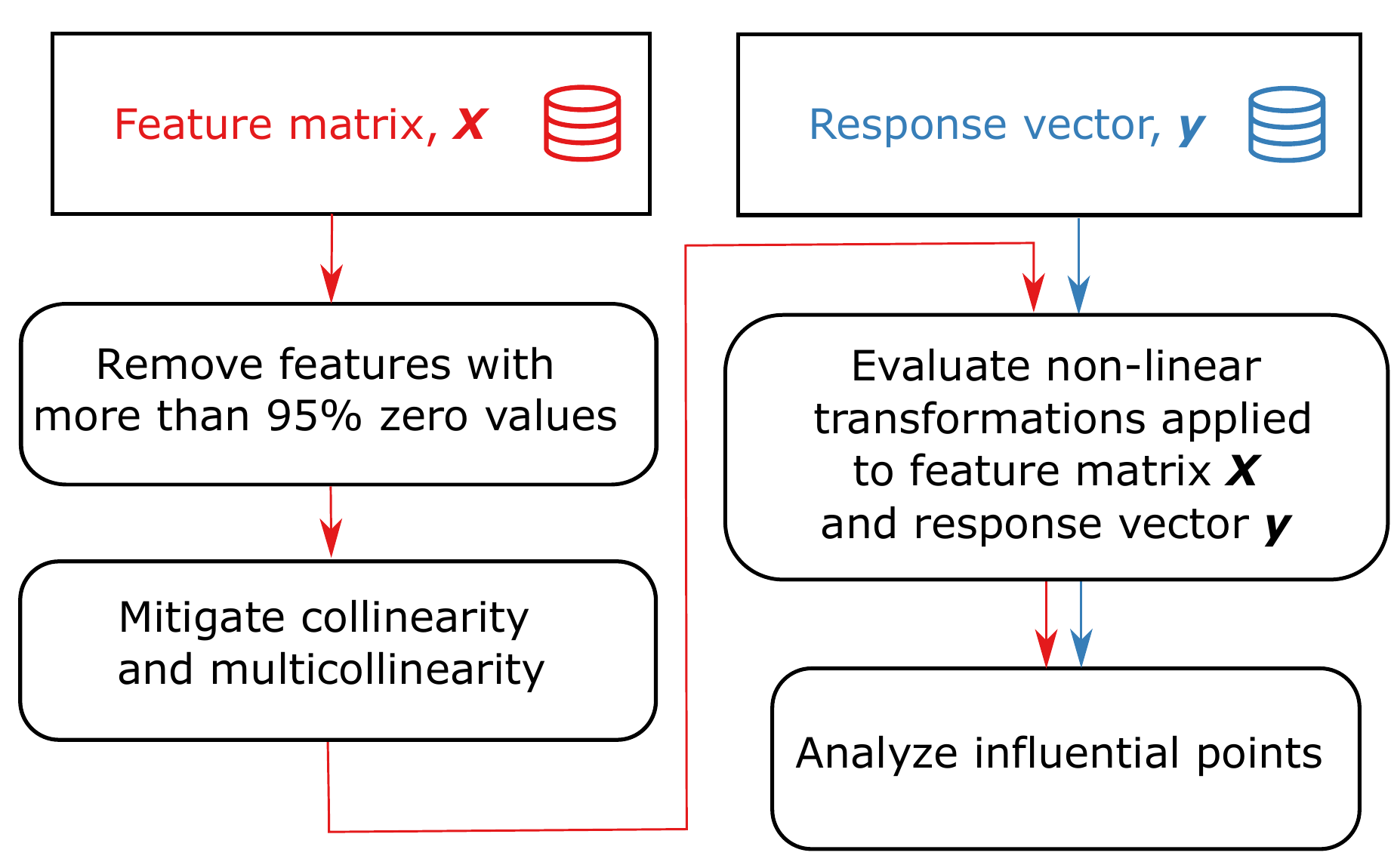}
	\caption{Schematic illustrating the workflow applied to the feature matrix $\bm{X}$ and response vector $\bm{y}$ to analyse potential data modelling problems.}
	\label{fig:data_preprocessing}
\end{figure}

Adopting the notation from Ref.~\cite[p.~5]{hastie2015statistical}, features are organised in a matrix $\bm{X} = (\bm{x}_1,~\dots~,\bm{x}_p)$, where the column~$\bm{x}_j$ corresponds to the feature $j~=~1,~\dots,~p$ and~$x_{ij}$ is the~$i$-th observation of the feature~$j$, for $i~=~1,~\dots,~n$, where~$n$ is the number of observations (charging pools). The response vector~$\bm{y}$ represents the energy consumption in kWh on charging pools during the year~2015, i.e.~$y_i$ is the energy consumption of the charging pool~$i$.

When a dataset is fit with a regression model, many problems may occur. Several steps, recommended in the literature~\cite{James_2014,Kuhn_2013,Afifi_2011}, were applied to the feature matrix~$\bm{X}$ and to the response vector~$\bm{y}$, to explore and to address potential problems. The sequence of steps is illustrated in Figure~\ref{fig:data_preprocessing}. 

First, uninformative features, i.e. those with more than~$95$\% of zero values, were excluded. Dependencies between features can cause serious problems when interpreting results obtained by regression methods, referred to as collinearity and multicollinearity. Despite the potential to bias the interpretation of results, these problems tend to be overlooked in the data analysis~\cite{hsu2015identifying}. We identified groups of features with absolute value of the Pearson correlation coefficient between each pair of features in the group greater than $0.95$~\cite[p.~47]{Kuhn_2013}. For each group, we chose one feature as a representative of the group. The selected feature was included, while other group members were excluded from further analysis. Identified groups and selected representative features are listed in Table~\ref{tab:correlated_features} of the SI file. To mitigate multicollinearity, we applied a procedure composed of two steps: in the first step, the values of the variance inflation factor (VIF)~\cite{James_2014} were calculated. In the second step, we identified feature with the maximum value of the VIF. If the maximum VIF was greater than or equal to value $10$, the corresponding feature was excluded from further analysis and the procedure was repeated, otherwise, the procedure was terminated. The list of excluded features is provided in Section~\ref{secSI:modelling_problems} of the SI file. These steps reduced the multicollinearity to the level recommended in the literature, while quantified not only by the values of the VIF but also by values of measures derived from eigenvalues of the correlation matrix~\cite[p.~252]{chatterjee2015regression}.

To explore whether a nonlinear function of $\bm{X}$ could explain response vector $\bm{y}$ better than the linear function, we transformed original features $j~=~1, \dots, p$, except binary features, using functions $\sqrt{\bm{x}_j}$, $\bm{x}_j^2$ and $\log(\bm{x}_j+1)$ and the feature matrix $\bm{X}$ was extended by transformed features (in this section, functions are applied to vectors element-wise). By combining the ordinary least squares (OLS) with 10-fold cross-validation~\cite{James_2014}, the extended feature matrix was fit to the response vector $\bm{y}$ and the mean squared error (MSE) was evaluated on the test data. From the results, we concluded that the basic non-linear transformations do not improve the fit sufficiently to compensate for additional model complexity and for making the model more likely to overfit the data. Hence, we do not apply any non-linear transformation to the feature matrix $\bm{X}$. Furthermore, we evaluated several transformations of the response vector: $\sqrt{\bm{y}}$, $\bm{y}^2$, $\log(\bm{y})$ and the Box-Cox transformation~\cite[p.~32]{Kuhn_2013}. By analysing the residual plots (see Figure~\ref{fig:residual_plots} of the SI file) we concluded that the transformation $\log(\bm{y})$ significantly improves the fit and we apply it in the regression analyses when fitting the energy consumption with the feature matrix, presented in Section~\ref{sec:data_analyses_methods}. 

Finally, we obtained $1259$ observations and $119$ features derived from geospatial datasets characterizing the vicinity of charging pools and $5$ features derived from EVnetNL dataset specifying the location and basic characteristics of charging pools.  
\subsection{Data analyses methods}
\label{subsec:data_analyses_methods}
This paper aims at explaining consumed energy on charging pools using features derived from the geospatial datasets. This task can be formulated as a regression problem combined with the statistical inference, indicating the statistical strength of features. From the geospatial datasets, we extracted a relatively large number of features. To facilitate the interpretation of results, features that have a higher potential to explain response variable shall be selected. We assume that from all features, only a relatively small number plays an important role. Hence, we use the Lasso method~\cite{Tibshirani96}, which combines the parameter fitting with the variable selection functionality.
\subsubsection{The Lasso method}
\label{subsec:lasso_method}
Considering the collection of $n$ samples $\{(\bm{x}_i, y_i) \}_{i=1}^{n}$, for some $\lambda \geq 0$ the Lasso method solves optimization problem 
\begin{equation} 
\underset{\beta_0, \beta}{\text{minimize}}  \left\{ \frac{1}{2n} \sum_{i=1}^{n} (y_i - \beta_0 - \sum_{j=1}^{p} x_{ij}\beta_j)^2 + \lambda \sum_{j=1}^{p} |\beta_j| \right\},
\label{eq_lasso}
\end{equation}
where the scalar $\beta_0$ (intercept) and vector $\beta$ (regression coefficients) are optimization variables. The first term corresponds to the least squares objective function and its role is to ensure a good fit between the linear regression model $\eta(\bm{x}_i) = \beta_0 + \sum_{j=1}^{p} x_{ij} \beta_j$ and response value $y_i$, while the second term regularizes the estimated values of regression coefficients in a way that leads to a variable selection (i.e. some regression coefficients are set to zero). The parameter $\lambda$ determines the trade-off between goodness of the fit and strength of the variable selection. Often, the factor $\frac{1}{2n}$ in (\ref{eq_lasso}) is replaced, with $1/2$ or $1$. Although this corresponds to a simple reparametrization of $\lambda$, the factor $\frac{1}{2n}$ makes $\lambda$ values comparable across different sample sizes, which is useful for cross-validation~\cite{hastie2015statistical}. The solution of problem~(\ref{eq_lasso}), $\hat{\beta_0}$ and $\hat{\beta}$, constitutes the estimate of the model parameters. 

The Lasso method tends to have some difficulties with the identification of relevant features on datasets with highly correlated features~\cite[p.~55]{hastie2015statistical}. The Elastic Net method, i.e. adding the term $\lambda_2 \sum_{j=1}^{p} |\beta_j|^2$ (with $\lambda_2 \geq 0$) to (\ref{eq_lasso}), may help to identify correlated features~\cite[p.~56]{hastie2015statistical}. In numerical experiments, the Elastic Net method was tested as well, however, it selected largely the same set of features as the Lasso method. 
\subsubsection{Model selection}
\label{sec:model_selection}
The parameter $\lambda$ in (\ref{eq_lasso}) controls the complexity of the model. A smaller value of $\lambda$ results in a larger number of non-zero regression coefficients and allows the model to adapt more closely to data, however, it can lead to overfitting. On the contrary, a larger value of $\lambda$ leads to a sparser and more interpretable model with the risk of preventing the Lasso from capturing the main signal in the data. Hence, the value of $\lambda$ should be carefully chosen. To estimate a suitable value of $\lambda$, we evaluated a range of values using log spacing. For each value, we used the $k$-fold cross-validation~\cite{hastie2015statistical} to evaluate the MSE of the model. Finally, we found the $\lambda^{CV}$ for which the minimum MSE was achieved and we selected the corresponding $\hat{\beta}_0^{CV}$ and $\hat{\beta}^{CV}$.
\subsubsection{Statistical inference}
\label{sec:statistical_inference}
Traditional methods, such as OLS, determine the statistical strength of features by evaluating $p$-values. The results obtained by the OLS regression on features selected by the Lasso method cannot be fully used in post-selection analysis as the exclusion of some features causes a bias~\cite[p.~155]{hastie2015statistical}. The adaptive nature of the Lasso method makes the problem of estimating $p$-values difficult--both conceptually and analytically. From the available approaches to the inference problem, we selected the bootstrap~\cite[p.~142]{hastie2015statistical}. The bootstrap was recommended as a suitable method for the assessment of the stability of selected regression coefficients in~\cite{hsu2015identifying}, even though there it has not been used in the analysis while taking a risk of presenting an unreliable set of significant coefficients. The bootstrap is a generic tool for assessing the statistical properties of complex estimators. First, the dataset is sampled. Second, the $k$-fold cross-validation is applied to each sample, to find $\lambda^{CV}$, $\hat{\beta}_0^{CV}$ and $\hat{\beta}^{CV}$. The distributions of $\hat{\beta}^{CV}$ coefficients obtained across all samples and the counts of samples resulting in coefficients $\hat{\beta}^{CV}$ equal to zero are used to assess the significance of features~\cite[p.~153]{hastie2015statistical}
\section{Results}
\label{sec:data_analyses_methods}
\subsection{Software libraries and settings}
\label{subsec:libraries_settings}
We prepared and modelled the data using the R language. We processed the GIS data with~\textit{sf},~\textit{raster} and~\textit{osmar} packages. The distribution of energy consumption was fitted with the~\textit{fitdistrplus} package. The Lasso method, including model selection, is implemented in the~\textit{glmnet} package. 
For the $k$-fold cross-validation we used $k$ = 10.
We considered the values $10^i$, for $i$ in the interval from $-4$ to $0$ in steps of $0.02$, when applying the cross-validation to explore the values of the parameter $\lambda$ in the Lasso method. When studying the stability of coefficients selected by the Lasso method, we used the bootstrap method with $10~000$ realisations.
\subsection{Metrics of consumed energy at charging pools}
\label{subsec:compare_energy_measures}
Charging pools differ in the maximum capacity and in the number of charging points (see panels A and B of Figure~\ref{fig:fig_1ab}), potentially leading to differences in the consumed energy. As illustrated in Figure~\ref{fig:distr_energy}A, charging pools with a higher number of charging points tend to have slightly higher energy consumption. Moreover, the number of cases when two transactions run on a charging pool in parallel is not negligible (see inset of Figure~\ref{fig:distr_energy}A). To analyze to what extent it is important to account for these effects, we define several simple metrics of the consumed energy at charging pools in Table~\ref{tab:corrResponses}.
    \begin{table}[]
        \centering
        \begin{tabular}{lr}
            Response variable characterizing a charging pool & $R^2$ \\
          \hline
            Consumed energy  & $0.380$ \\ 
            Consumed energy per charging point & $0.367$ \\
            Maximum energy consumed at a charging point & $0.360$ \\
            Consumed energy per one unit of charging capacity & $0.333$ \\
          \hline
         \end{tabular}
        \caption{Response vectors characterizing the energy consumed at a charging pool. The coefficient of determination, $R^2$, was obtained by fitting the logarithmic transformation of the response vector (see Figure~\ref{fig:residual_plots} of the SI file) with the feature matrix by applying the ordinary least squares method.}
        \label{tab:corrResponses}
    \end{table}
To gain some basic understanding, whether some measures can be better explained with the feature matrix, we ran the OLS regression. In Table~\ref{tab:corrResponses} we report the values of the coefficient of determination, $R^2$. The highest $R^2$ we obtained for the consumed energy at a charging pool. High similarity in $R^2$ values we attribute to the high correlations between response vectors. The Pearson correlation coefficient for all pairs of response vectors ranges from $0.86$ to $0.98$. Hence, for further analyses, we chose the consumed energy at a charging pool as the response vector~$\bm{y}$.
\subsection{Distribution of consumed energy over charging pools}
\label{subsec:energy_distribution}
\begin{figure*}
    \centering
    \includegraphics[width=0.75\textwidth]{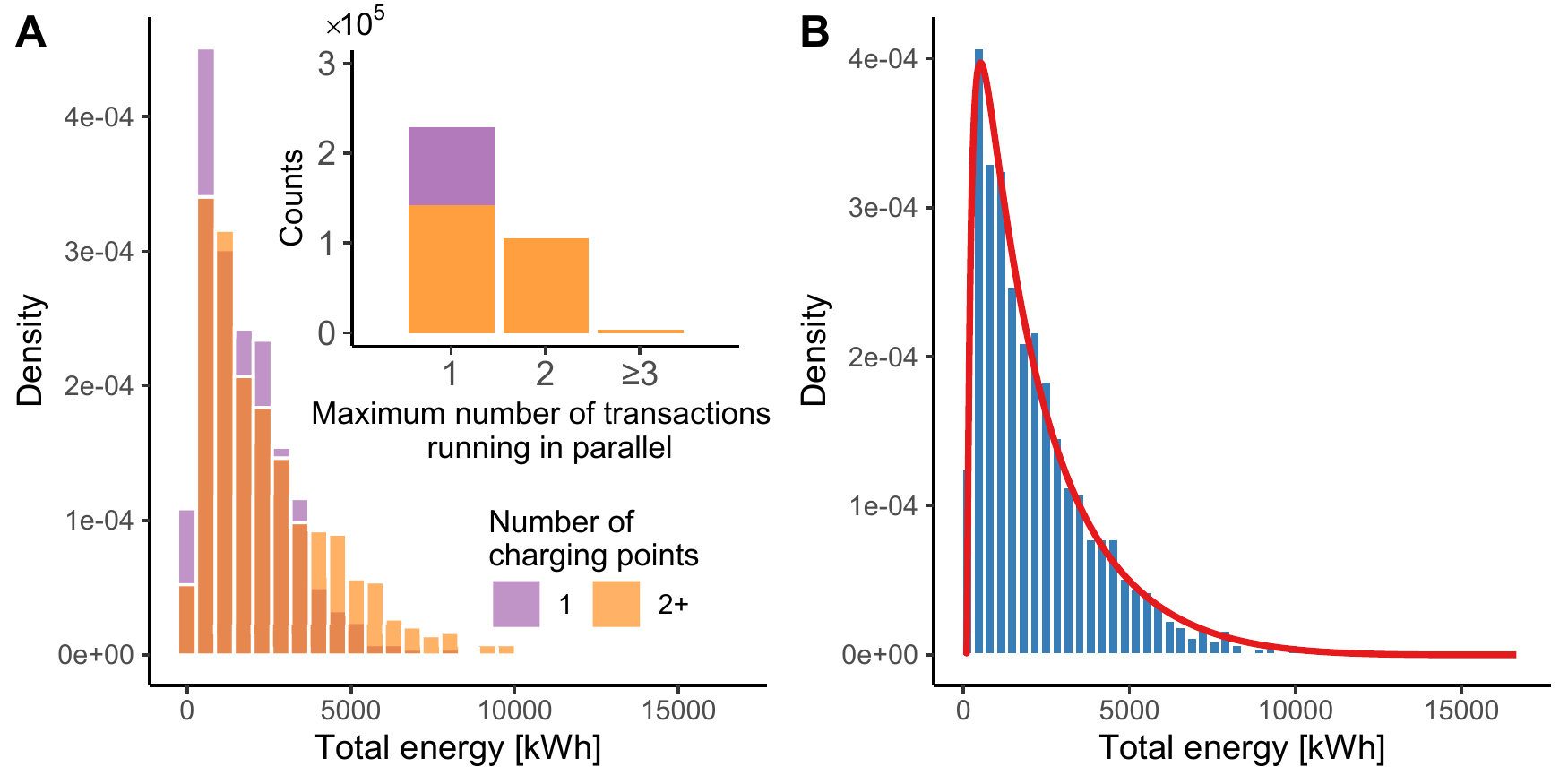}
    \caption{The empirical probability distribution of energy consumed at charging pools. \textbf{A}~We show two separate probability distributions, one for charging pools with only one charging point and one for charging pools with more than one charging point. The inset shows the stacked bar plot of the number of EVs charged in parallel with other EVs on a charging pool. \textbf{B}~The~empirical probability distribution of the total energy consumed at charging pools. The red line represents the fit obtained using Eq.~(\ref{eq:betaDistrCustom}).}
    \label{fig:distr_energy}
\end{figure*}
The density histogram indicates that the consumed energy at charging pools is rather heterogeneous and positively skewed (see Figure~\ref{fig:distr_energy}B). We observe high consumption on a small number of charging pools and small consumption is observed for a large group of charging pools. To model the distribution of energy consumed at charging pools, we considered original data and five transformations ($\bm{y}^2$, $\bm{y}^3$, $\sqrt{\bm{y}}$, $\sqrt[3]{\bm{y}}$ and $log(\bm{y})$) and parametrized density functions of three known probabilistic distributions (Weibull, beta and gamma). We used the Kolmogorov-Smirnov goodness of fit test together with the inspection of the P-P and the Q-Q plots~\cite{warren2010application} to conclude that the transformation $\sqrt[3]{\bm{y}}$ combined with the beta distribution provides the best fit to the data. The fitting procedure is detailed in Section~\ref{secSI:energyFitting} of the SI file. The functional form of the probability density function is derived in Appendix~\ref{sec:appendix} and takes the form
\begin{equation}
                f(y,\alpha,\beta) = \frac{\Big(\frac{\sqrt[3]{y}-y_{min}}{y_{max}-y_{min}}\Big)^{\alpha-1} \Big(1-\frac{\sqrt[3]{y}-y_{min}}{y_{max}-y_{min}}\Big)^{\beta-1}}{B(\alpha, \beta)2(y_{max}-y_{min}) y^{\frac{2}{3}}}.
    \label{eq:betaDistrCustom}
\end{equation}
The symbol $B(\alpha, \beta)$ denotes the beta function and the estimates of parameters take the following values: $\alpha = 2.58$ , $\beta = 4.53$, $y_{min} = 91.55$ kWh and $y_{max} = 16~649.40$ kWh. 
\subsection{Explaining the energy consumption from other charging pool performance indicators}
\label{subsec:energy_simple_models}
\begin{table}[]
\centering
\begin{tabular}{lrrrrr}
  \hline
 Model & $\hat{k}$ & $R^2$ & mean & stdev & cv\\ 
  \hline
  $\bm{y} = k\bm{n}$           & $8.10$   & $0.90$ & $8.69$   & $3.65$   & $0.42$\\ 
  $\bm{y} = k\bm{t}$           & $895.55$ & $0.59$ & $897.02$ & $764.74$ & $0.85$\\
  $\bm{y} = k\bm{p}$           & $632.67$ & $0.56$ & $678.46$ & $589.20$ & $0.87$\\ 
  $\bm{y} = k(\bm{t \circ p})$ & $245.25$ & $0.59$ & $269.05$ & $228.06$ & $0.85$\\ 
  $\bm{y} = k(\bm{n \circ p})$ & $2.49$   & $0.95$ & $2.52$   & $0.63$   & $0.25$\\ 
  $\bm{y} = k(\bm{n \circ t})$ & $3.25$   & $0.94$ & $3.45$   & $0.92$   & $0.27$\\ 
  \hline
\end{tabular}
  \caption{Simple regression models of the energy consumed at charging pools. In columns are presented the estimates of the regression coefficient $\hat{k}$; the corresponding $R^2$ value obtained by the OLS regression; the mean value of the quantity that is represented by the regression coefficient $k$ calculated over charging pools (mean); the corresponding standard deviation (stdev) and the coefficient of variation (cv) calculated from the mean and the standard deviation. The symbol $\circ$ denotes the component-wise multiplication (the Hadamard product) of vectors.}
  \label{tab:simple_models}
\end{table}
We gained interesting insights by analyzing the relationship between energy consumption and other charging pool indicators constituting the energy consumption. The energy consumption at a charging pool $i$ can be decomposed into the product of three other indicators, i.e.
\begin{equation}
y_i = n_i t_i p_i,
\label{eq:energy_pool_integral}
\end{equation}
where $n_i$ is the number of charging transactions taking place at a charging pool $i$, $t_i$ is the average charging time per transaction at a charging pool $i$ and $p_i$ is the average charging power at a charging pool $i$. We organized these quantities for all charging pools as vectors $\bm{n}$, $\bm{t}$ and $\bm{p}$. To assess the role of these three factors, in the heterogeneity of the consumed energy across charging pools, we explored six models presented in Table~\ref{tab:simple_models}. Models are based on Eq.~(\ref{eq:energy_pool_integral}), where one indicator or product of two indicators, represented by the regression coefficient $k$, is considered invariant across charging pools. 

We obtained the estimate $\hat{k}$ of the coefficient $k$ from the data by using the OLS method. Among the models that explain the consumed energy from one indicator, the highest value of $R^2$ is obtained for the model explaining the consumed energy from the number of charging transactions. Similarly, models explaining energy consumption from pairs of indicators that include the number of transactions have high values of $R^2$. Hence, the major factor associated with the heterogeneity in the consumed energy across charging pools is the number of transactions. The fluctuations in the charging patterns (i.e. average charging time and average charging power) play a much smaller role.

In Table~\ref{tab:simple_models}, we calculated, mean, standard deviation and the coefficient of variation over all charging pools for quantities which are represented by the coefficient $k$. For example, in the model $\bm{y} = k\bm{n}$, the coefficient $k$ that replaces in Eq.~(\ref{eq:energy_pool_integral}) the expression $t_i p_i$ can be interpreted as the average of consumed energy per transaction. Hence, the average consumed energy per transaction is $8.69$ kWh, the charging time per transaction takes on average $2.52$ hours and the average power reaches $3.45$ kW. These numbers indicate that the majority of charged EVs are plug-in hybrids. The charging capacity, which is for the majority of charging points around $11$ kW (see Figure~\ref{fig:fig_1ab}), is significantly underused. The values of the coefficient of variation are high when the number of transactions is included in the analysed quantity, confirming that the largest variance is associated with the number of transactions.

As the number of charging transactions is closely associated with the consumed energy at charging pools, it is crucial to understand the way EV drivers decide which charging opportunity they choose. We hypothesize that some exogenous factors, characterizing the environment surrounding the charging pools affect the consumed energy at charging pools and we explore it in the next section. 
\subsection{Explaining the energy consumption from geospatial data}
\label{sec:energy_consumption_from_urban_data}
\begin{figure*}
	\centering
	\includegraphics[width=0.9\textwidth]{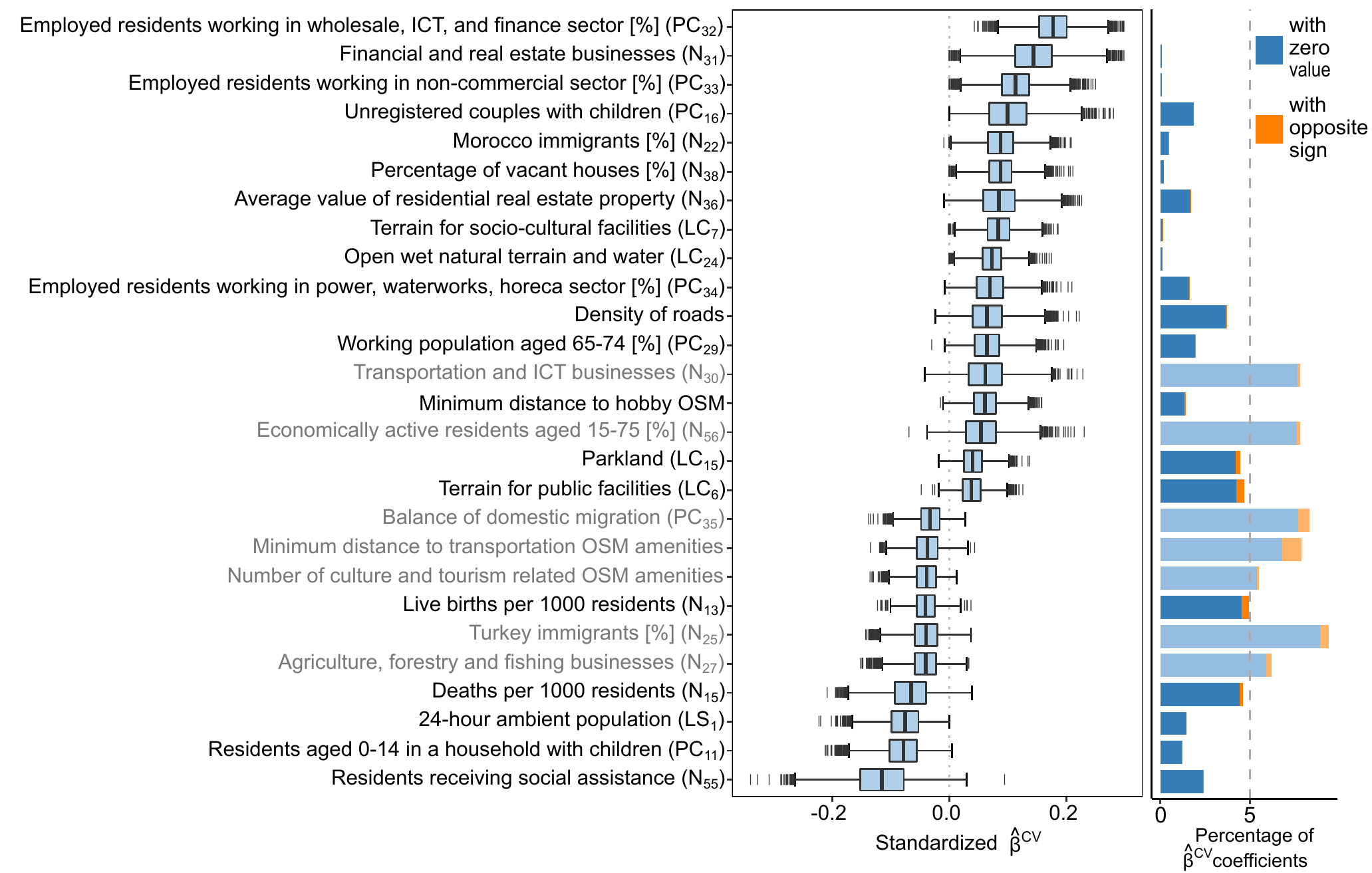}
	\caption{The empirical distributions of standardized regression coefficients obtained by the Lasso method and the $10$-fold cross-validation applied to $10~000$ samples of bootstrapped data. We show only features where the value of the regression coefficient was set to zero in less than $10\%$ of the samples. Coefficients are descendingly ordered, from the largest to the smallest median value. The left panel presents the Tukey's box plot of coefficients. On the right, the stacked bar plot shows the percentage of samples when the regression coefficient $\hat{\beta}^{CV}$ was set to zero and the number of samples where it reached the opposite sign as the sign of the median. We consider as significant those features where the number of samples with zero coefficient value is less than $5\%$ and the number of samples with opposite sign is small. The dashed line indicates the $5\%$ threshold value. Features that are not considered significant we display using faded colours. Full descriptions of coefficients can be found in tables~\ref{tabSI:Population_cores_attr}~-~\ref{tabSI:landscan} of the SI file by using the code in brackets.}
	\label{fig:coef_imp_lasso_boot}
\end{figure*}

We applied the methodology described in Section~\ref{subsec:data_analyses_methods} to the feature matrix derived from geospatial datasets, first without considering the features derived from the EVnetNL dataset (see Section~\ref{secSI:features}), and to the $log$-transformation of the response vector $\bm{y}$, representing the consumed energy at charging pools. To facilitate the mutual comparison of regression coefficients, we standardized each coefficient by multiplying it with the standard deviation of the bootstrap sample of the predictor. Thus, the standardized coefficient can be considered as an estimate of the change in the response variable, when the feature increases by one standard deviation. The larger is the absolute value of the median of standardized regression coefficient samples, the stronger is the feature's potential to influence the response variable~[p.~372]\cite{siegel2016practical}. Hence, the absolute value of the median of bootstrap realisations can be considered as a measure of the feature strengths. The different signs of regression coefficients, across bootstrap realisations, can be attributed to the low significance or to the simultaneous selection of correlated features~\cite[p.~144]{hastie2015statistical}. Hence, the more consistent are the values of standardized regression coefficient across bootstrap samples, the more significant is the feature corresponding to a regression coefficient. As a rule of thumb, we consider as significant those features where the number of bootstrap samples with zero coefficient value is less than $5\%$ and the number of samples with the opposite coefficient sign to the sign of the majority of the sample is negligible. To provide a broader view on results, we show in Figure~\ref{fig:coef_imp_lasso_boot} the empirical distributions of standardized regression coefficients that reached the value of zero in maximum $10\%$ of the bootstrap realisations. The statistical inference is sometimes omitted in energy studies and only a single run of a variable selection method is evaluated~\cite{hsu2015identifying, ma2016estimation}. The inspection of Tukey's boxplots justifies the use of the bootstrap or in more general, the use of a statistical inference method. The majority of presented coefficients obtained a zero value in few samples or exceptionally the opposite sign, that could lead to omitting these coefficients or interpreting them in a wrong way if evaluating only a single run of the Lasso method.

We show the regression coefficients in descending order of median value in Figure~\ref{fig:coef_imp_lasso_boot}. To clarify the results, we organized significant features with the positive sign of the median in four groups:
\begin{itemize}
    \item \textbf{Physical environment(+):} \textit{terrain for social and cultural facilities} ($LC_7$); \textit{open wet natural terrain and water} ($LC_{24}$); \textit{density of roads}; \textit{parkland} ($LC_{15}$); \textit{terrain for public facilities} ($LC_6$).
    \item \textbf{Population(+):} \textit{employed residents working in the wholesale, ICT and finance sector [\%]} ($PC_{32}$); \textit{employed residents working in non-commercial sector [\%]} ($PC_{33}$); \textit{unregistered couples with children} ($PC_{16}$); \textit{Morocco immigrants [\%]} ($N_{22}$); \textit{employed residents working in power, waterworks, horeca (hotel/ restaurant/caf\'{e}) sector [\%]} ($PC_{34}$); \textit{working population aged 65 - 74 [\%]} ($PC_{29}$).
    \item \textbf{Services and businesses(+):} \textit{financial and real estate businesses} ($N_{31}$); \textit{minimum distance to hobby OSM}.
    \item \textbf{Buildings(+):} \textit{percentage of vacant houses} [\%] ($N_{38}$); \textit{average value of residential real estate property} ($N_{36}$).
\end{itemize}

Similarly, we organized significant features with the negative value of the median
into a group:
\begin{itemize}
    \item \textbf{Population(-):} \textit{residents receiving social assistance} ($N_{55}$); \textit{residents aged 0 - 14 living in a household with children} ($PC_{11}$); \textit{24-hour ambient population} ($LS_{1}$); \textit{ deaths per 1000 residents } ($N_{15}$); \textit{live births per 1000 residents} ($N_{13}$).
\end{itemize}

The largest number of significant features we found in the population group, which is partly because this group contains most of the features. Many significant features point to a single factor. The largest group of features, $PC_{32}$, $PC_{29}$, $N_{36}$, ($N_{55}$), indicate that high (low) income and wealth (poverty) are positively (negatively) linked with the amount of consumed energy at charging pools. Most likely, this is due to the high prices of EVs, making them more affordable for better-situated residents and businesses. Probably for the same reasons, some significant features ($PC_{11}$, $N_{13}$) are linked with children or youth and to the elderly or retired population ($N_{15}$), i.e. social groups that are typically less wealthy. The high \textit{percentage of vacant houses} ($N_{38}$) and the high \textit{average value of residential real estate property} ($N_{36}$) are associated with high energy consumption at charging pools. It could correspond to newly built and yet not entirely inhabited areas, with a higher standard of living expressed in higher real estate values. Note, if a feature representing a minimum distance to an object has a positive coefficient, then the energy consumption increases with increasing the distance from a given object and vice versa. Hence, the proximity of \textit{hobby related points of interest} is negatively associated with the energy consumption at charging pools. Furthermore, the \textit{density of roads} is also among features that are positively linked with the energy consumption at charging pools, indicating that good access to charging pools contributes to energy consumption.

We included to the feature matrix some features (\textit{number of charging points}, \textit{maximum power}, \textit{latitude} and \textit{longitude} and the \textit{rollout strategy}) derived from the EVnetNL dataset (see Section~\ref{secSI:features}) and repeated the analysis. The results are shown in Figure~\ref{figSI:energPoolFeatures} of the SI file. In general, significant features are similar as in Figure~\ref{fig:coef_imp_lasso_boot}; however, the number of significant features is slightly smaller. The significance of some features from groups Physical environment(+) and Population(-) was reduced. All EVnetNL features are significant, while the \textit{maximum charging capacity}, the \textit{number of charging points} and \textit{longitude} have high strength, indicating that parameters of charging pools potentially influence the energy consumption. We attribute the reduction in the number of significant features to the replacement of some features by EVnetNL features. For example, previously significant feature, \textit{open wet natural terrain and water} ($LC_{24}$), seems to be expressed via the \textit{longitude}. The negative influence of \textit{longitude} can be explained by the geography of the Netherlands, whereas the western part of the country is more urbanized and we find here the majority of large Dutch cities. At the same time, there is a lot of surface water as the western part of the country is largely situated below the sea level.
\subsection{Influence of the rollout strategy on the energy consumption}
\label{subsec:rollout_strategy_effects}
\begin{figure*}
    \centering
    \includegraphics[width=0.8\textwidth]{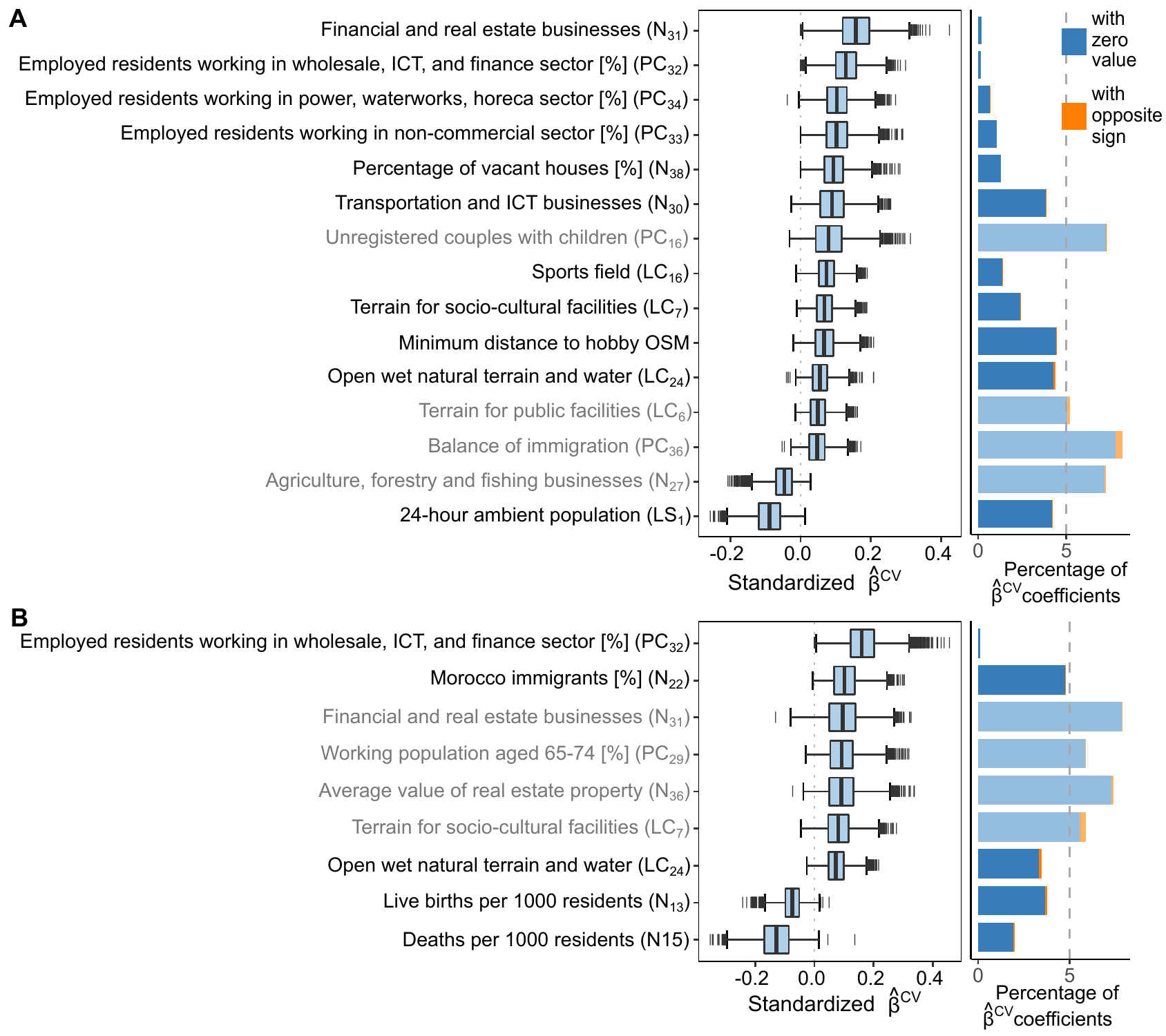}
    \caption{The empirical distributions of standardized regression coefficients obtained by the Lasso method combined with the $10$-fold cross-validation applied to $10~000$ samples of bootstrapped data. \textbf{A}~Strategic rollout of charging pools. \textbf{B}~Demand-driven rollout of charging pools. We show only features where the value of the regression coefficient was set to zero in less than $10\%$ of the samples. Regression coefficients are descendingly ordered from the largest to the smallest median value. The left panel presents the Tukey's box plot of coefficients. On the right, the stacked bar plot shows the percentage of samples where the regression coefficient $\hat{\beta}^{CV}$ was set to zero and the number of samples where it reached the opposite sign as the sign of the median. We consider as significant those features where the number of samples with zero coefficient value is less than $5\%$ and the number of samples with opposite sign is small. The dashed line indicates the $5\%$ threshold value. Features that are not considered significant we display using faded colours. Full descriptions of coefficients can be found in tables~\ref{tabSI:Population_cores_attr}~-~\ref{tabSI:landscan} of the SI file by using the code in brackets.} 
    \label{fig:coef_imp_rollout} 
\end{figure*}
The majority of charging pools has been located using one out of two (strategic or demand-driven) rollout strategies~\cite{Helmus_2018}. The strategically located charging pools are placed near public facilities, where the EV charging is intuitively expected. The demand-driven charging pools are built upon the request from EV users, typically near to their homes. In this section, we investigate whether the rollout strategy makes a difference in factors associated with energy consumption. Information about the rollout strategy is available in the EVnetNL dataset and we used it to split charging pools into two groups. We applied the Lasso method separately to each group considering the feature matrix without the features derived from the EVnetNL dataset. The selected features in Figure~\ref{fig:coef_imp_rollout} coincide to a large extent with factors selected for the complete dataset (see Figure~\ref{fig:coef_imp_lasso_boot}), however, now we can observe differentiation of factors according to the location strategy. 

The energy consumed at strategically located charging pools, (Figure~\ref{fig:coef_imp_rollout}A), is positively linked to the working sector of residents and the physical environment, i.e. to certain types of venues adjacent to the charging pool. Working sectors of residents represented by $PC_{32}$, $PC_{33}$ and $PC_{34}$ indicate the prevalent businesses in municipalities, positively associated with energy consumption. Moreover, selected features for strategical rollout refer to some businesses and some locations (\textit{sports fields}, \textit{socio-cultural facilities}) that could be associated with occasional charging.

For the charging pools with the demand-driven rollout, (Figure~\ref{fig:coef_imp_rollout}B), the negative coefficients of \textit{deaths per 1000 residents} and \textit{live births per 1000 residents} indicate that areas with higher natality and mortality are negatively linked with the energy consumption, pointing out to areas inhabited by socially weaker groups of specific age categories.

We have tested several other stratifications, e.g. based on the \textit{number of charging points}, the proportion of the residential area within the charging pool's buffer, the administrative division of the Netherlands into provinces and the number of residents of municipalities. Except for the last criterion, we obtained only a very small number of selected features. 
Dividing charging pools into two groups based on the municipality population, considering a threshold of~$50~000$ residents, we obtained approximately the same size of groups.
Interestingly, we observe that charging pools located in municipalities with more than~$50~000$ residents consume more energy on average by~$48$\% than charging pools in the other group of municipalities.
We find the higher number of significant features, linked to municipality population characteristics, \textit{financial and real estate businesses} and the physical environment, for the charging pools located in municipalities with a smaller population (see Figure~\ref{figSI:energyCities} in the SI file).
\section{Conclusions and discussion}
We analyzed the explainability of consumed energy at charging pools from several points of view. Main conclusions derived from the data analysis are the following:
\begin{itemize}
\item The energy consumption can be satisfactorily modelled by a transformed beta distribution.
\item The number of charging transactions is the driving factor among the characteristics constituting energy consumption.
\item The economic prosperity appears to be behind a large group of regression variables selected for the mathematical description of the relationship between energy consumption and locational factors derived from available geospatial datasets. For example, residents and businesses with high (low) income, situated in the charging pool vicinity, are linked to a positive (negative) impact on energy consumption. Similarly, charging pools located close to expensive newly built housing show higher energy consumption. 
The western part of the Netherlands with four major large cities is positively linked to energy consumption as well.
Considering the standardized values of regression coefficients, certain working sectors of municipalities' residents and the \textit{number of financial and real estate businesses} have a large positive impact on the energy consumption. The largest adverse impact have \textit{residents receiving social assistance}.

\item The stratification of charging pools by the rollout strategy leads to the split of selected regression coefficients. Business types, working sector of residents and public venues in the proximity are linked to higher consumption of energy at charging pools deployed strategically. Population characteristics, e.g. \textit{live births} and \textit{deaths per 1000 residents} are linked to the energy consumption at charging pools placed based on the demand.
\end{itemize}

Our results extend the knowledge base about the energy consumption at charging pools and provide an evidence for a relationship between energy consumption and the characteristics of the vicinity of charging pools opening several possibilities for future applications.

The methodology and our findings can be used to fine-tune rollout strategies for deployment of charging pools. A rollout strategy optimized for a specific group of charging pools, e.g. charging pools used for work-charging, can be enhanced by applying the presented methodology to this specific group of charging pools and identified characteristics can be used to select locations of new charging pools in a way which corresponds better to energy constraints. Information on selected regression variables can be used to build prediction models in a more targeted way. The presented results are applicable to the Netherlands, however, the proposed methodology can be used also elsewhere. 

The selected regression coefficients can be associated with three different spatial scales. Some describe close vicinity of charging pools, e.g. \textit{the number of financial and real estate businesses} ($N_{31}$), others are attributed to the municipality, e.g. features derived from the Population cores datasets such as the \textit{percentage of employed residents in the municipality working in a non-commercial sector} ($PC_{33}$). The last group of regression coefficients has the potential to characterize the location of a charging pool at the country level, e.g. \textit{latitude} and \textit{longitude}~(see Figure~\ref{figSI:energPoolFeatures} in the SI file). Hence, while properly considering the spatial level of selected regression coefficients, a hierarchical or a customized rollout strategy, considering specific geographic scales, could be designed.

Apart from enhancing the strategies for charging pools deployment, our study can help to improve energy demand models for power grid capacity planning by better considering the energy demand at charging pools and the proposed methodology could be also applied to other service systems, e.g. to stations for shared electric cars, scooters or bicycles.
\subsection{Limitations and further research}
\label{sec:limitations_and_further_research}
Several important limitations are inherited from the used statistical methods. Apart from notorious limitation ``correlation does not imply causation'', which requires careful consideration of all results, we wish to point out limitation to the interpretability of our results due to the presence of multicollinearity in the input data. To obtain statistically stable results, we reduced the level of multicollinearity to the level recommended in the literature. However, the removal of some features hampers the interpretation of our results. Similarly, it is likely that some relevant data representing important determinants of energy consumption are missing in the analyses, e.g. mobility behaviour of the population or visitation patterns of venues located in the vicinity of charging pools, as we were not able to collect them. Further research could focus on a more complex characterization of the usage of charging pools, specific groups of charging pools e.g. determined based on similarities in usage patterns, or exploration of possibilities to extend the proposed methodology and findings to other geographic areas.

\appendix
\section{Appendix: Fitting the energy consumption with the transformed beta probability density function}
\label{sec:appendix}

Considering the Kolmogorov-Smirnov goodness-of-fit test together with the inspection of the P-P and the Q-Q plots~\cite{warren2010application}, we concluded that the most satisfactory fit of the consumed energy at charging pools is obtained by transforming the data with the function $g(\bm{y}) = \sqrt[3]{\bm{y}}$ (the function is applied to the vector element-wise) and using the beta distribution (see section~\ref{secSI:energyFitting} of the SI file). The beta distribution is defined on the interval $\langle0, 1\rangle$. After the rescaling, the beta distribution is suitable to represent a random variable between a minimum value $y_{min}$ and a maximum value $y_{max}$. Hence, with the beta distribution, we model the random variable
\begin{equation}
Z = \frac{\sqrt[3]{Y} - y_{min}}{y_{max} - y_{min}},
\label{eq:transformed_RV}
\end{equation}
where the random variable $Y$ is modelling the energy consumption. Using Eq.~(\ref{eq:transformed_RV}), we establish the relation between the distribution function of $Y$ and $Z$ as
\begin{align}
F_Y(y) & = P(Y < y) = \nonumber \\
       &   P((Z(y_{max}-y_{min}) + y_{min})^3 < y) =  \\
       &   P\Big(Z < \frac{\sqrt[3]{y}-y_{min}}{y_{max}-y_{min}}\Big) = F_Z\Big(\frac{\sqrt[3]{y}-y_{min}}{y_{max}-y_{min}}\Big) \nonumber.
\label{eq:distributionFunctionDeduction}
\end{align}
Consequently, the density function characterizing the random variable $Y$ is
\begin{align}
f_Y(y) & = \big[ F_Y(y) \big]' = \bigg[ F_Z\Big(\frac{\sqrt[3]{y}-y_{min}}{y_{max}-y_{min}}\Big) \bigg]' = \nonumber \\ 
& F'_Z\Big(\frac{\sqrt[3]{y}-y_{min}}{y_{max}-y_{min}}\Big) \bigg[ \frac{\sqrt[3]{y}-y_{min}}{y_{max}-y_{min}} \bigg]' = \\
& f_Z\Big(\frac{\sqrt[3]{y}-y_{min}}{y_{max}-y_{min}}\Big) \frac{1}{3(y_{max}-y_{min}) y^{\frac{2}{3} } }.\nonumber
\label{eq:densityFunctionDeduction}
\end{align}
Since $f_Z(\bm{z})$ is the probability density function of the beta distribution, we get the following density function for the energy consumption

\begin{equation}
f_Y(y, \alpha, \beta) = \frac{\Big(\frac{\sqrt[3]{y}-y_{min}}{y_{max}-y_{min}}\Big)^{\alpha-1} \Big(1-\frac{\sqrt[3]{y}-y_{min}}{y_{max}-y_{min}}\Big)^{\beta-1}}{B(\alpha, \beta)3(y_{max}-y_{min}) y^{\frac{2}{3}} },
\label{eq:MyBetaDensityFunction}
\end{equation}
where $B(\alpha, \beta)$ is the beta function with parameters $\alpha$ and $\beta$.

\printcredits
\section*{Acknowledgements}
We thank \v{L}udmila J\'{a}no\v{s}\'{i}kov\'{a} and Luca Lena Jansen for valuable comments and suggestions.
\bibliographystyle{cas-model2-names}


\newpage
\setcounter{section}{0}
\setcounter{figure}{0}
\setcounter{table}{0}
\renewcommand{\thetable}{S\arabic{table}}
\renewcommand{\thesection}{S\arabic{section}}
\renewcommand{\thefigure}{S\arabic{figure}}

\vspace{5 mm}
\section*{\centering{\Large{Supplementary Information: Explaining the distribution of energy consumption at slow charging infrastructure for electric vehicles from socio-economic data}}}
\vspace{10 mm}

\section{Datasets}
\label{secSI:datasets}
We characterised the vicinity of charging infrastructure with publicly available datasets on population, socio-economic data, land use, energy consumption, traffic flows, and points of interest (see Table~\ref{tabSI:GISdata}). Each dataset contains several attributes, some of which are replicated on other datasets. To avoid duplicates and dependencies, we kept only the attribute from the dataset with the highest resolution and excluded attributes that are the sum of other attributes. The initial list of selected attributes for each dataset is shown in Tables~\ref{tabSI:Population_cores_attr}-~\ref{tabSI:openchargemap}.

\begin{longtable}[h]{p{.15\textwidth} |p{.55\textwidth} |p{0.17\textwidth}| p{.05\textwidth}} 
	\arrayrulecolor{lightgray}
	\toprule 	
	Name & Description & File format/resolution & Source\\
	\hline
	Population cores & Population cores are continuous spatial units with at least~25 houses or~50 residents containing various demographic and socio-economic data at municipality level spatial resolution. & shapefile/ heterogeneous polygons & \cite{popcores} \\
	\hline
	Neighbourhoods & Neighbourhoods contain data on population, living, households, companies, social security, motor vehicles and services. Neighbourhoods are the smallest spatial areas used for the collection of statistical data in the Netherlands. & shapefile/ neighbourhoods & \cite{neighbpop} \\
	\hline
	Land use & The dataset contains digital geometry and the corresponding description of land use classes, \eg roads, buildings, recreational areas and indoor and outdoor water. The boundaries of land use areas are based on the digital topographical base Top10NL. & shapefile/custom & \cite{lccbs} \\ 
	\hline
	Energy consumption & The dataset describes the annual consumption of natural gas and electricity of residential houses and industrial buildings together with the number of buildings where the consumption was measured. & shapefile/ neighbourhoods & \cite{energyatlas} \\
	\hline
	Liveability & The dataset quantifies a quality of living based on more than 100 features divided into five categories: houses, socio-economic background of residents, services, safety and environment. The dataset is created and maintained by the Dutch Ministry of the Interior and Kingdom Relations.
	& shapefile/ neighbourhoods & \cite{liveability} \\
	\hline
	Traffic flows &  The dataset consists of the average flow of cars, buses and trucks for every road segment in the Dutch road network. & shapefile/road segments & \cite{traffic_flows}\\
	\hline
	LandScan (ambient population) & The ambient population is the 24-hour average of population in a given area. The  dataset includes information about daily mobility, \eg industrial areas with low residential population but high population during the day may have high ambient population. & raster/30$\times$30 arc seconds & \cite{landscan}\\ 
	\hline
	OpenStreetMap & The dataset consists of points of interest laying inside $2\times 2\text{ km}^2$ polygons centred at EVnetNL charging pools. The points of interest are organised in 15~categories. & point data & \cite{osm} \\
	\hline
	OplaadPaalen & The dataset contains the position of $5~794$ charging stations located in the Netherlands. & point data & \cite{opp} \\
	\hline
	OpenChargeMap & The dataset contains the position and some characteristics of $5~458$ charging stations located in the Netherlands. & point data & \cite{ocm} \\
	\hline
	\hline
	\caption{Overview of the geospatial datasets that are publicly available for the territory of the Netherlands and we identified them as relevant for the analysis.}
	\label{tabSI:GISdata}
\end{longtable}
\subsection{Population cores}
\label{secSI:population_cores}
\begin{longtable}[h]{ p{.10\textwidth} |p{.75\textwidth} |p{.10\textwidth}} 
	\toprule
	Identifier & Description & Data type\\
	\hline
	$PC_{1}$ & Average age of the population within the core area by January 1, 2011. & average \\ 
	\hline
	$PC_{2}$ & Number of individuals aged 15 - 24 in one-person households. & count \\ 
	\hline
	$PC_{3}$ & Number of individuals aged 25 - 44 in one-person households. & count \\ 
	\hline
	$PC_{4}$ & Number of individuals aged 45 - 64 in one-person households. & count \\ 
	\hline
	$PC_{5}$ & Number of individuals aged 65 or older in one-person households. & count \\ 
	\hline
	$PC_{6}$ & Number of individuals aged 0 - 14 in multi-person households without biological or adopted children, or stepchildren.  & count \\ 
	\hline
	$PC_{7}$ & Number of individuals aged 15 - 24 in multi-person households without biological or adopted children, or stepchildren.  & count \\ 
	\hline
	$PC_{8}$ & Number of individuals aged 25 - 44 in multi-person households without biological or adopted children, or stepchildren.  & count\\ 
	\hline
	$PC_{9}$ & Number of individuals aged 45 - 64 in multi-person households without biological or adopted children, or stepchildren. & count\\ 
	\hline
	$PC_{10}$ & Number of individuals aged 65 or older in multi-person households without biological or adopted children, or stepchildren.  & count\\ 
	\hline
	$PC_{11}$ & Number of individuals aged 0 - 14 in multi-person households with biological or adopted children, or stepchildren.  & count \\ 
	\hline
	$PC_{12}$ & Number of individuals aged 15 - 24 in multi-person households with biological or adopted children, or stepchildren.  & count \\ 
	\hline
	$PC_{13}$ & Number of individuals aged 25 - 44 in multi-person households with biological or adopted children, or stepchildren.  & count \\ 
	\hline
	$PC_{14}$ & Number of individuals aged 45 - 64 in multi-person households with biological or adopted children, or stepchildren.  & count \\ 
	\hline
	$PC_{15}$ & Number of individuals aged 65 or more in multi-person households with biological or adopted children, or stepchildren.  & count \\ 
	\hline
	$PC_{16}$ & Number of individuals living as an unmarried couple or couple without registered civil partnership  with biological or adopted children, or stepchildren.  & count \\ 
	\hline
	$PC_{17}$ & Number of individuals living as an unmarried couple or couple without registered civil partnership without children that belong to a household of two people. & count \\
	\hline 
	$PC_{18}$ & Number of individuals living as a married or couple with registered civil partnership with children. & count \\ 
	\hline
	$PC_{19}$ & Number of individuals living as a married or couple with registered civil partnership without children that belong to a household of two people. & count \\ 
	\hline
	$PC_{20}$ & Number of individuals in private households of one parent with at least one child living at home. & count \\ 
	\hline
	$PC_{21}$ & Number of individuals living in a private household but not in a relationship.  & count \\ 
	\hline
	$PC_{22}$ &  Number of individuals inhabiting institutional properties, such as nursing homes, elderly and children's homes, group homes, rehabilitation centres and prisons. & count \\ 
	\hline
	$PC_{23}$ & Number of individuals whose parents were both born in the Netherlands, regardless of their homeland. & count \\ 
	\hline
	$PC_{24}$ & Number of individuals with at least one parent born in Europe (excluding Turkey), North America, Oceania, Indonesia or Japan. & count \\ 
	\hline
	$PC_{25}$ & Number of individuals with at least one parent born in Africa, Latin America, Asia (excluding Indonesia and Japan) or Turkey. & count \\ 
	\hline
	$PC_{26}$ & Percentage of the working population $25$ - $44$ years old. & percentage \\
	\hline 
	$PC_{27}$ & Percentage of the working population $45$ - $54$ years old. & percentage \\ 
	\hline
	$PC_{28}$ & Percentage of the working population $55$ - $64$ years old. & percentage \\ 
	\hline
	$PC_{29}$ & Percentage of the working population $65$ - $74$ years old. & percentage \\ 
	\hline
	$PC_{30}$ & Percentage of the working population working in agriculture, forestry and fishing. & percentage \\ 
	\hline
	$PC_{31}$ & Percentage of the working population working in mining, manufacturing and construction. & percentage \\ 
	\hline
	$PC_{32}$ & Percentage of the working population employed in commercial services corresponding to the categories: wholesale and retail, transportation and storage, information and communication, financial services, hire and selling of real estate, lease of movable goods and other business services, veterinary services. & percentage \\ 
	\hline
	$PC_{33}$ & Percentage of the working population engaged in non-commercial services corresponding to the categories: public administration and public services, education, health and welfare, culture, sports and recreation, other services, households as employers, extraterritorial organisations. & percentage \\ 
	\hline
	$PC_{34}$ & Percentage of the working population engaged in one of the following categories: power, waterworks and waste management, hotel, restaurant and cafes, specialist business services, except veterinary services (Categories not counted in $PC_{30}$-$PC_{33}$). & percentage \\ 
	\hline
	$PC_{35}$ & Number of individuals moving into the geographical area of the core minus the number of individuals moving elsewhere in the Netherlands in the period $1$ January $2001$ - $1$ January $2011$ & count \\ 
	\hline
	$PC_{36}$ & Net number of immigrants into the core (the number of immigrants moving into the core minus the number of individuals who emigrated in the period $1$ January $2001$ - $1$ January $2011$). & count \\ 
	\hline
	$PC_{37}$ & Number of households with two individuals. & count \\ 
	\hline
	$PC_{38}$ & Number of households with three individuals. & count \\
	\hline 
	$PC_{39}$ & Number of households with four individuals. & count \\
	\hline 
	$PC_{40}$ & Number of households with five individuals. & count \\ 
	\hline
	$PC_{41}$ & Number of households with six or more individuals. & count \\ 
	\hline
	$PC_{42}$ & Average house value in EUR. & average \\ 
	\hline
	$PC_{43}$ & Average value of an owner-occupied house in EUR. & average \\ 
	\hline
	$PC_{44}$ & Average value of rental housing in EUR. & average \\ 
	\hline
	$PC_{45}$ & Number of recreational properties within the core area. & count \\ 
	\hline
	\hline
	\caption{List of attributes selected from the dataset \emph{Population cores}.}
	\label{tabSI:Population_cores_attr}
\end{longtable}

\newpage

\subsection{Neighbourhoods}
\label{secSI:Neighbourhoods}

\begin{longtable}[h]{ p{.10\textwidth} |p{.75\textwidth} |p{.10\textwidth}} 
	\toprule
	Identifier & Description & Data type\\
	\hline
	$N_{1}$ & The percentage of addresses in the neighbourhood with the most frequent zip code. The most frequent zip code is the zip code assigned to the largest number of addresses in the neighbourhood. This attribute is organized in six categories by the percentage of addresses sharing the same zip code. Category one: more than $90\%$ of addresses, category two: $81$ - $90$\%, category three: $71$ - $80$\%, category four: $61$ - $70$\% , category five: $51$ - $60$\%, category six: less than $50$\% of addresses share the same zip code.& categorical \\ 
	\hline
	$N_{2}$ & The average of address densities for a neighbourhood. For each address within a neighbourhood, a circular area with a radius of $1$ kilometre centred at the address is considered. First, the density of addresses within each circular area is determined. Second, the average of address densities for a neighbourhood is calculated as the average of address densities considering all circular areas. This value is used as an estimate of the level of urbanity of an area. & average \\
	\hline
	$N_{3}$ & Urban class of the neighbourhood, based on the density of properties (five classes). & categorical \\
	\hline
	$N_{4}$ & Total population. & count \\
	\hline
	$N_{5}$ & Percentage of population aged $15$ - $24$ years. & percentage \\
	\hline
	$N_{6}$ & Percentage of population aged $25$ - $44$ years. & percentage \\
	\hline
	$N_{7}$ & Percentage of population aged $45$ - $64$ years. & percentage \\
	\hline
	$N_{8}$ & Percentage of population aged $65$ years or older. & percentage \\
	\hline
	$N_{9}$ & Percentage of residents who are neither married nor in a registered civil partnership. & percentage \\
	\hline
	$N_{10}$ & Percentage of residents who are married or in a registered civil partnership. & percentage \\
	\hline
	$N_{11}$ & Percentage of widowed residents.  & percentage \\
	\hline
	$N_{12}$ & Number of live births in $2015$. & count \\
	\hline
	$N_{13}$ & Number of live births in $2015$, per thousand residents. & per thousand \\
	\hline
	$N_{14}$ & Number of all deaths in $2015$. & count \\
	\hline
	$N_{15}$ & Number of deaths in $2015$, per thousand residents. & per thousand \\
	\hline
	$N_{16}$ & Number of households. & count \\
	\hline
	$N_{17}$ & Percentage  of private households with one person. & percentage \\
	\hline
	$N_{18}$ & Percentage of private households that have more than one resident but no children. & percentage \\
	\hline
	$N_{19}$ & Average household size. & average \\
	\hline
	$N_{20}$ & Percentage of immigrants with a western origin (Europe (excluding Turkey), North America and Oceania or Indonesia or Japan), relative to the entire population. & percentage \\
	\hline
	$N_{21}$ & Percentage of immigrants with a non-Western origin, relative to the entire population. These immigrants originate from Turkey or the continents of Africa, Latin America and Asia (excluding Indonesia and Japan), relative to the entire population. & percentage \\
	\hline
	$N_{22}$ & Percentage of immigrants with Moroccan origin relative to the entire population. & percentage \\
	\hline
	$N_{23}$ & Percentage of immigrants with origin from (former) Netherlands Antilles and Aruba, relative to the entire population. & percentage \\
	\hline
	$N_{24}$ & Percentage of immigrants with origin from Suriname, relative to the entire population.  & percentage \\
	\hline
	$N_{25}$ & Percentage of immigrants with origin from Turkey, relative to the entire population. & percentage \\
	\hline
	$N_{26}$ & Percentage of immigrants with another non-western origin, relative to the entire population, $N_{26}=N_{21} - (N_{22} + N_{23} + N_{24} + N_{25})$. & percentage \\
	\hline
	$N_{27}$ & Number of agriculture, forestry and fishing businesses. & count\\ 
	\hline
	$N_{28}$ & Number of industry and energy businesses. & count\\
	\hline
	$N_{29}$ & Number of trade, hotels and restaurants. & count\\
	\hline
	$N_{30}$ & Number of transportation, information and communication businesses. & count\\
	\hline
	$N_{31}$ & Number of financial services, real estate businesses. & count\\
	\hline
	$N_{32}$ & Number of business services. & count\\
	\hline
	$N_{33}$ & Number of cultural, recreational, and other services not included in $N_{27}$-$N_{32}$. & count\\
	\hline
	$N_{34}$ & Number of business establishments.& count \\
	\hline		
	$N_{35}$ & Total number of residential properties with at least one residential function and possibly one or more other usage functions. & count \\
	\hline
	$N_{36}$ & Average value of residential real estate property in thousands of EUR. & average \\
	\hline
	$N_{37}$ & Number of multi-family housing as a percentage of the total housing stock.  & percentage \\
	\hline
	$N_{38}$ & Percentage of vacant houses. & percentage \\
	\hline
	$N_{39}$ & Percentage of owner-occupied properties. & percentage \\
	\hline
	$N_{40}$ & Percentage of rental properties. & percentage \\
	\hline
	$N_{41}$ & Percentage of rental properties owned by a housing organization. & percentage \\
	\hline
	$N_{42}$ & Percentage of rental properties owned by other landlords than housing organizations (e.g. properties owned by a person). & percentage \\
	\hline
	$N_{43}$ & Percentage of houses built in the year $2000$ or later, relative to the total number of houses. & percentage \\
	\hline
	$N_{44}$ & Number of individuals receiving an income. & count\\
	\hline
	$N_{45}$ & Average income in thousand EUR per an individual receiving income. & average\\ 
	\hline
	$N_{46}$ & Average income in thousand EUR per resident. & average\\
	\hline
	$N_{47}$ & Percentage of individuals in private households belonging to the nationwide $40$\% with the lowest personal income. & percentage\\
	\hline
	$N_{48}$ & Percentage of individuals in private households belonging to the nationwide $20$\% with the highest personal income. & percentage\\
	\hline
	$N_{49}$ & Percentage of private households belonging to the nationwide $40$\% households with the lowest household income. & percentage\\
	\hline
	$N_{50}$ & Percentage of private households belonging to the nationwide 20\% households with the highest household income. & percentage\\
	\hline
	$N_{51}$ & Percentage of households with low purchasing power. & percentage\\
	\hline
	$N_{52}$ & Percentage of households below or around social minimum, except student households. The \emph{social minimum} is a government set benchmark for the cost of a decent life in the region. & percentage\\
	\hline
	$N_{53}$ & Number of individuals who receive a disability benefit under the Occupational Disability Insurance Act (WAO), Self-employed Persons Occupational Disability Insurance Act (WAZ), work and income according to the Labour Capacity Act (WIA), Young Disabled Persons Act (Wajong). & count\\
	\hline
	$N_{54}$ & Number of individuals receiving benefits under the Unemployment Insurance Act. & count\\
	\hline
	$N_{55}$ & Number of individuals receiving social assistance benefits under the Employment and Assistance Act or the Participation Act. & count\\
	\hline
	$N_{56}$ & Percentage of economically active individuals aged $15$ - $75$ years. & percentage\\
	\hline
	$N_{57}$ & Number of individuals receiving a basic pension from the central government under the General Old Age Pensions Act. & count\\
	\hline
	$N_{58}$ & Number of motor vehicles for road passenger transport, excluding mopeds and motorcycles, with up to nine seats (including the driver). & count\\
	\hline
	$N_{59}$ & Number of passenger cars per household. & count\\
	\hline
	$N_{60}$ & Number of vans, lorries, tractors (motor vehicle equipped to tow trailers), selected vehicles (commercial vehicles for selected purposes such as fire trucks, cleaning cars, tow trucks) and buses. & count\\
	\hline
	$N_{61}$ & Number of motorcycles, scooters, motor carriers and motor wheelchairs with a motorcycle registration. & count\\
	\hline
	$N_{62}$ & Number of cars aged six years and older. & count\\
	\hline
	$N_{63}$ & Number of petrol cars. & count\\
	\hline
	\hline
	\caption{List of attributes selected from the dataset \emph{Neighbourhoods}.}
	\label{tabSI:Neighbourhoods}
\end{longtable}

\subsection{Land use}
\label{secSI:land_use}

\begin{longtable}[h]{ p{.10\textwidth} |p{.75\textwidth} |p{.10\textwidth}} 
	\toprule
	Identifier & Description & Data type\\
	\hline
	$LC_{1}$ & Railway. & category \\
	$LC_{2}$ & Road. & category \\
	$LC_{3}$ & Airport. & category \\
	$LC_{4}$ & Residential area. & category \\
	$LC_{5}$ & Terrain for retail and catering industry. & category \\
	$LC_{6}$ & Terrain for public facilities. & category \\
	$LC_{7}$ & Terrain for social and cultural facilities. & category \\
	$LC_{8}$ & Business area. & category \\
	$LC_{9}$ & Dumping ground. & category \\
	$LC_{10}$ & Car wreck storages and scrap yards. & category \\
	$LC_{11}$ & Cemetery. & category \\
	$LC_{12}$ & Mines, oil and gas fields. & category \\
	$LC_{13}$ & Building site. & category \\
	$LC_{14}$ & Semi paved other terrains (dikes not covered with grass, undeveloped land that is not considered as a building site, and railway lines no longer in use). & category \\
	$LC_{15}$ & Parkland. & category \\
	$LC_{16}$ & Sports field. & category \\
	$LC_{17}$ & Terrain for non-commercial ornamental and vegetable cultivation (allotments and school gardens). & category \\
	$LC_{18}$ & Recreational terrain used for one day (one-day camping terrain, zoo and safari parks, amusement parks, marinas and open-air museums). & category \\
	$LC_{19}$ & Residential ground in use for a multi-day recreational stay (camping, bungalow parks and youth hostels). & category \\
	$LC_{20}$ & Land for greenhouse horticulture. & category \\
	$LC_{21}$ & Other agricultural lands. & category \\
	$LC_{22}$ & Forest. & category \\
	$LC_{23}$ & Open dry natural terrain. & category \\ 
	$LC_{24}$ & Water (open wet natural terrain and water). & category \\
	$LC_{25}$ & Water used for recreational purposes (water in golf courses and parks, rowing lanes and recreational pools). & category \\
	\hline
	\caption{Categories of the polygons in the \emph{Land Use} dataset.}
	\label{tabSI:land_use}
\end{longtable}

\subsection{Energy consumption}
\label{secSI:energy_consumption}

\begin{longtable}[h]{ p{.10\textwidth} |p{.75\textwidth} |p{.10\textwidth}} 
	\toprule
	Identifier & Description & Data type\\
	\hline
	$EC_{1}$ & Mean gas consumption of residential properties [$m^3$].& average \\
	$EC_{2}$ & Annual gas consumption of residential properties [$m^3$]. & count \\
	$EC_{3}$ & Number of residential properties where gas consumption was measured. & count \\
	$EC_{4}$ & Mean electricity consumption of residential properties [kWh].& average \\
	$EC_{5}$ & Annual consumption of electric energy of residential properties [kWh]. & count \\
	$EC_{6}$ & Number of residential properties where electric energy was measured. & count \\
	$EC_{7}$ & Mean gas consumption of companies [$m^3$].& average \\
	$EC_{8}$ & Annual gas consumption of companies [$m^3$]. & count \\
	$EC_{9}$ & Number of companies where gas consumption was measured. & count \\
	$EC_{10}$ & Mean consumption of electric energy by companies [kWh].& average \\
	$EC_{11}$ & Annual electric energy consumption by companies [kWh]. & count \\
	$EC_{12}$ & Number of companies where electric energy consumption was measured. & count \\
	\hline
	\caption{List of attributes selected from the dataset \emph{Energy Consumption}.}
	\label{tabSI:energy_consumption}
\end{longtable}

\subsection{Liveability}
\label{secSI:liveability}

\begin{longtable}[h]{ p{.10\textwidth} |p{.75\textwidth} |p{.10\textwidth}} 
	\toprule
	Identifier & Description & Data type\\
	\hline
	$L_{1}$ & Liveability index 2016 - subcategory houses, comprehending factors like type, built year and ownership of houses (deviation from the national average). &  numeric\\
	\hline
	$L_{2}$ & Liveability index 2016 - subcategory socio-economic background of residents, comprehending types of families, migration background, unemployment rate and similar population characteristics (deviation from the national average). & numeric \\
	\hline
	$L_{3}$ & Liveability index 2016 - subcategory services, consisting mostly of availability of walking distance to various services (deviation from the national average). & numeric \\
	\hline
	$L_{4}$ & Liveability index 2016 - subcategory safety, evaluating crime rates of different types (deviation from the national average).  & numeric \\
	\hline
	$L_{5}$ & Liveability index 2016 - subcategory environment, embracing vicinity of the neighbourhood, \eg forests, factories, wind turbines, highways etc. (deviation from the national average). & numeric \\
	\hline
	$L_{6}$ & Liveability index 2016 - composite index quantifying the quality of life. & categorical ordinal \\
	\hline
	\caption{List of attributes selected from the dataset \emph{Liveability}.}
	\label{tabSI:liveability}
\end{longtable}

\subsection{Traffic flows on road segments}
\label{secSI:traffic_flows}

\begin{longtable}[h]{ p{.10\textwidth} |p{.75\textwidth} |p{.10\textwidth}} 
	\toprule
	Identifier & Description & Data type\\
	\hline
	$TF_{1}$ & Number of cars per hour (7--19 hr). & count \\
	$TF_{2}$ & Number of cars per hour (19--23 hr). & count \\
	$TF_{3}$ & Number of cars per hour (23--7 hr). & count \\
	$TF_{4}$ & Number of buses per hour (7--19 hr). & count \\
	$TF_{5}$ & Number of buses per hour (19--23 hr). & count \\
	$TF_{6}$ & Number of buses per hour (23--7 hr). & count \\
	$TF_{7}$ & Number of trucks per hour (7--19 hr). & count \\
	$TF_{8}$ & Number of trucks per hour (19--23 hr). & count \\
	$TF_{9}$ & Number of trucks per hour (23--7 hr). & count \\
	\hline
	\caption{List of attributes associated with the road segments of the dataset \emph{Traffic Flows}.}
	\label{tabSI:traffic_flows}
\end{longtable}		

\newpage

\subsection{Landscan}
\label{secSI:landscan}

\begin{longtable}[h]{ p{.10\textwidth} |p{.75\textwidth} |p{.10\textwidth}} 
	\toprule
	Identifier & Description & Data type\\
	\hline
	$LS_{1}$ & Ambient population (population averaged over 24 hours) distribution. & count \\
	\hline
	\caption{List of attributes extracted from the \emph{Landscan} dataset.}
	\label{tabSI:landscan}
\end{longtable}

\subsection{OpenStreetMap}
\label{secSI:openstreetmap}

To find out if some points of interest are correlated with the energy usage of charging pools, we extracted the \emph{Points of Interest} from the OpenStreetMap (OSM), which are originally organised in $593$ categories for the Netherlands. As the majority of categories have just a few points, we re-aggregated the data into $15$ new categories based on common characteristics. The column Number of aggregated OSM categories in Table~\ref{tabSI:openstreetmap} shows the number of original categories aggregated.

\begin{longtable}[h]{p{.10\textwidth}p{.30\textwidth}>{\centering}p{.40\textwidth}p{.10\textwidth}} 
	\toprule
	Identifier & Description & Number of aggregated OSM categories & Data type\\
	\hline
	$OSM_{1}$ & Health. & 39 & point \\
	$OSM_{2}$ & Entertainment. & 147 & point \\
	$OSM_{3}$ & Culture and tourism. & 26 & point \\
	$OSM_{4}$ & Finance. & 10 & point \\
	$OSM_{5}$ & Fashion. & 49 & point \\
	$OSM_{6}$ & Food. & 42 & point \\
	$OSM_{7}$ & Transportation. & 36 & point \\
	$OSM_{8}$ & Work. & 26 & point \\
	$OSM_{9}$ & Household. & 76 & point \\
	$OSM_{10}$ & Education. & 17 & point \\
	$OSM_{11}$ & Public facilities. & 29 & point \\
	$OSM_{12}$ & Hobby. & 37 & point \\
	$OSM_{13}$ & Sport. & 39 & point \\
	$OSM_{14}$ & Accommodation. & 11 & point \\
	$OSM_{15}$ & Family. & 9 & point \\
	\hline
	\caption{List of categories compiled from the set of Points of Interest extracted from the OSM of the Netherlands.}
	\label{tabSI:openstreetmap}
\end{longtable}		

\subsection{Charging Pools 2015}
\label{secSI:charging_pools_2015}

To estimate the positions of all available charging pools in the Netherlands by the end of the year 2015, we compiled the Charging Pools 2015 dataset from the EVnetNL, OpenChargeMap~\cite{ocm} and OplaadPalen~\cite{opp} datasets using the date when a charging station was added to the dataset. We started with positions of all charging pools in the EVnetNL database, and from OpenChargeMap and OplaadPalen we added, one by one, the charging stations that were distant $50$ meters or more from charging pools already included. This process resulted in $8~366$ charging pools.

\begin{longtable}[h]{ p{.10\textwidth} p{.75\textwidth} p{.10\textwidth}} 
	\toprule
	Identifier & Description & Data type\\
	\hline
	$OCM_{1}$ & Geographic coordinates. & point \\
	\hline
	\caption{Attribute of the Charging Pools 2015 dataset.}
	\label{tabSI:openchargemap}
\end{longtable}		
\section{Materials and Methods}
\label{secSI:materials_and_methods}
\subsection{Handling missing values}
\label{subsecSI:handling}
\begin{figure}
	\centering
	\includegraphics[width=0.7\textwidth]{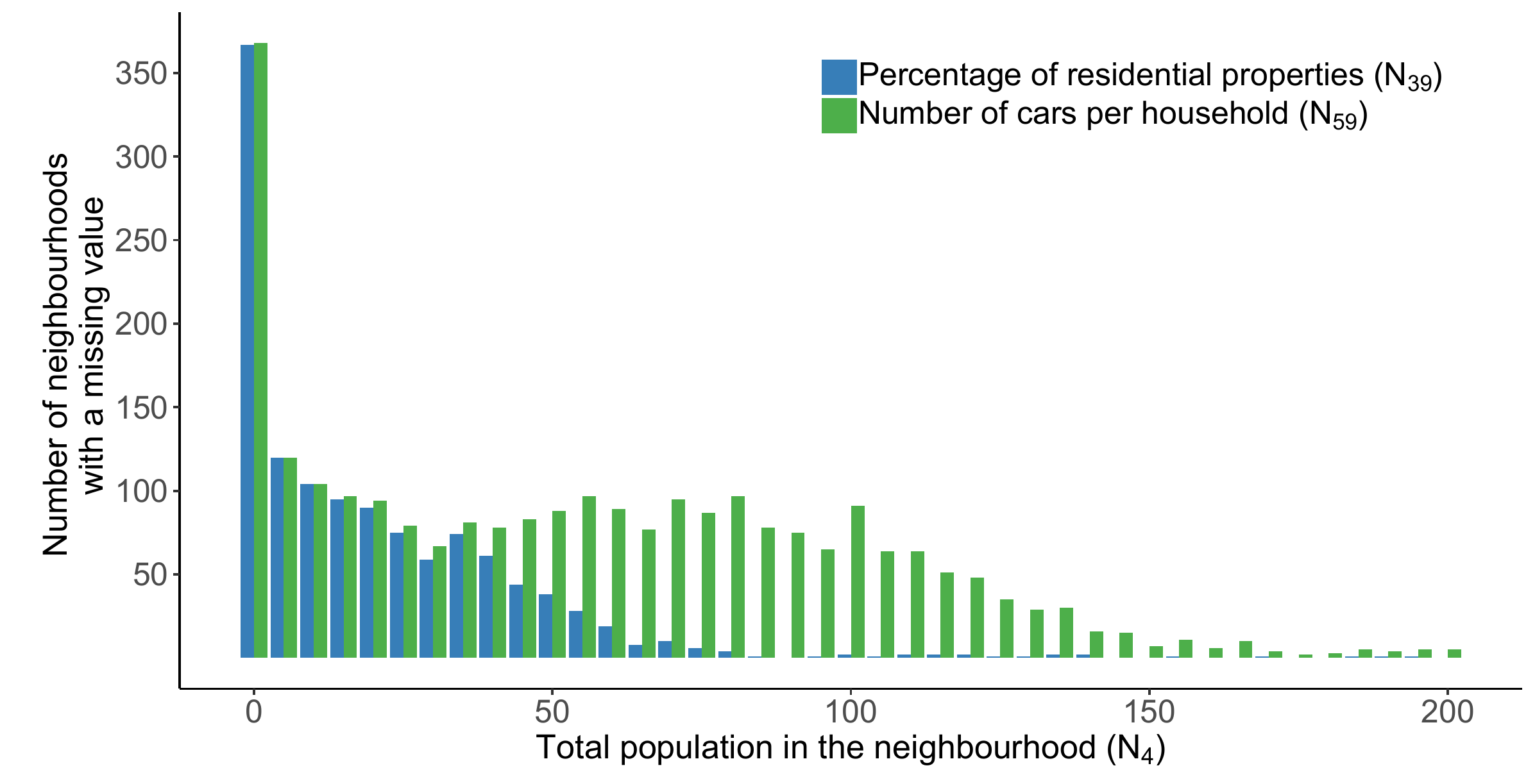}
	\caption{Illustration of the dependence between the number of missing values of attributes in Neighbourhoods dataset and the \textit{total population} $(N_{4})$. For both selected attributes, the \textit{percentage of owner-occupied properties} ($N_{39}$) and the \textit{number of cars per household} ($N_{59}$), we observe that the number of missing values in data is decreasing with the population.}
	\label{fig:missing_val_inhabitants}
\end{figure}

We handled the missing values of the attributes of geometric objects in the geospatial datasets in two steps. First, we imputed missing entries by the values that we derived from other attributes:

\begin{itemize}
	\item We estimated the \textit{number of individuals receiving a basic pension from the central government under the General Old Age Pensions Act} ($N_{57}$) from the \textit{percentage of population aged 65 years or older} ($N_{8}$) and the \textit{total population} ($N_{4}$). 
	
	\item We imputed the \textit{average household size} ($N_{19}$) by the ratio of the \textit{total population} ($N_{4}$) to the \textit{number of households} ($N_{16}$).  
	
	\item We imputed the \textit{number of passenger cars per household} ($N_{59}$) by dividing the \textit{number of motor vehicles for road passenger transport, excluding mopeds and motorcycles, with up to nine seats (including the driver)} ($N_{58}$) by the \textit{number of households} ($N_{16}$).
\end{itemize}

Second, we observed that attributes in the Neighbourhoods and Population cores datasets were often not available in areas with a low population (for examples see Figure~\ref{fig:missing_val_inhabitants}). Hence, we applied the following set of \textbf{if-then} rules to impute zero values or lowest possible values for missing values of attributes in areas with low population or low level of human activities:

\begin{enumerate}
	\item[Rule 1:]
	\begin{algorithmic}
		\IF{\textit{total population} ($N_{4}$) is zero,}
		\STATE set to the value of zero the missing values of attributes: \textit{average income in thousand EUR per resident} ($N_{46}$); \textit{number of individuals who receive a disability benefit under the Occupational Disability Insurance Act (WAO), Self-employed Persons Occupational Disability Insurance Act (WAZ), work and income according to the Labour Capacity Act (WIA), Young Disabled Persons Act (Wajong)} ($N_{53}$); \textit{number of individuals receiving benefits under the Unemployment Insurance Act} ($N_{54}$); \textit{number of individuals receiving social assistance benefits under the Employment and Assistance Act or the Participation Act} ($N_{55}$).
		\ENDIF    
	\end{algorithmic}
	\item[Rule 2:]
	\begin{algorithmic}
		\IF{\textit{total number of residential properties with at least one residential function and possibly one or more other usage functions} ($N_{35}$) is zero,}
		\STATE set to the value of zero the missing values of attributes: \textit{percentage of owner-occupied properties} ($N_{39}$); \textit{percentage of rental properties} ($N_{40}$); \textit{percentage of rental properties owned by s housing organization} ($N_{41}$); \textit{percentage of rental properties owned by other land lords than housing organization (e.g. properties owned by a person)} ($N_{42}$); \textit{percentage of houses built in the year 2000 and later} ($N_{43}$); \textit{percentage of vacant houses} ($N_{38}$); \textit{number of multi-family housing as a percentage of the total housing stock} ($N_{37}$); \textit{average value of residential real estate property in thousands of EUR} ($N_{36}$).
		\ENDIF    
	\end{algorithmic}
	\item[Rule 3:]
	\begin{algorithmic}
		\IF{ \textit{number of individuals receiving an income} ($N_{44}$) is zero,}
		\STATE set to the value of zero the missing values of attributes: \textit{average income in thousand EUR per an individual receiving income} ($N_{45}$); \textit{percentage of individuals in private households belonging to the nationwide 40\% with lowest personal income} ($N_{47}$); \textit{percentage of individuals in private households belonging to the nationwide 20\% with highest personal income} ($N_{48}$).
		\ENDIF    
	\end{algorithmic}
	
	\item[Rule 4:]
	\begin{algorithmic}
		\IF{ \textit{number of households} ($N_{16}$) is zero,}
		\STATE set to the value of zero the missing values of attributes: \textit{percentage of private households belonging to the nationwide 40\% households with the lowest household income} ($N_{49}$); \textit{percentage of private households belonging to the rural 20\% households with highest household income} ($N_{50}$); \textit{percentage of households below or around social minimum, except student households }($N_{52}$).
		\ENDIF    
	\end{algorithmic}
	
	\item[Rule 5:]
	\begin{algorithmic}
		\IF{ the number of households in the population core ($PC_{2} + PC_{3} + PC_{4} + PC_{5} + PC_{37} + PC_{38} + PC_{39} + PC_{40}+ PC_{41}$) is zero,}
		\STATE set to the value of zero the missing values attributes: \textit{average house value in EUR} ($PC_{42}$); \textit{average value of an owner-occupied house in EUR} ($PC_{43}$); \textit{average value of rental housing in EUR} ($PC_{44}$).
		\ENDIF    
	\end{algorithmic}
	
	\item[Rule 6:]
	\begin{algorithmic}
		\IF{ \textit{urbanity class of the neighbourhood based on the density of properties} ($N_{3}$), is missing,}
		\STATE set value of $N_{3}$ to class $5$ (meaning non-urban areas, i.e. the class corresponding to the category with the lowest urbanity class). 
		\ENDIF    
	\end{algorithmic}
	
	\item[Rule 7:]
	\begin{algorithmic}
		\IF{ \textit{percentage of addresses in the neighbourhood with the most frequent zip code} ($N_{1}$) is missing,}
		\STATE set value of $N_{1}$ to class $6$ (meaning that less than 50\% of addresses within the neighbourhood share the same zip code). 
		\ENDIF    
	\end{algorithmic}
	
	\item[Rule 8:]
	\begin{algorithmic}
		\IF{ the number of measurement points for the electricity or gas ($EC_{3}$, $EC_{6}$, $EC_{9}$, or $EC_{12}$) is less than or equal to five,}
		\STATE set the corresponding missing values of the electricity or gas consumption ($EC_{1}$, $EC_{2}$, $EC_{4}$, $EC_{5}$, $EC_{7}$, $EC_{8}$, $EC_{10}$, $EC_{11}$) to the value of zero. 
		\ENDIF    
	\end{algorithmic}
\end{enumerate}
\subsection{Feature engineering}
\label{secSI:features}
To model the vicinity of charging pools, we used circular buffers of radius $r$ meters centred at the positions of charging stations. We intersect buffers with geometric objects (points, polylines, polygons) in the geospatial dataset. To estimate the attribute values for buffers, we assumed a uniform spatial distribution of attribute values within polygons. Each attribute from datasets Population Cores ($45$~attributes), Neighbourhoods ($63$~attributes), Energy Consumption ($12$~attributes), Liveability ($6$~attributes) and Landscan ($1$~attribute) listed in the Tables~\ref{tabSI:Population_cores_attr}--\ref{tabSI:landscan} was used to derive one feature, resulting in $127$ features.

Furthermore, we created $25$ features by calculating the area taken by each land use category (listed in Table~\ref{tabSI:land_use}) within each buffer area. We created two features from each class of points of interest in the OpenStreetMap dataset (Table~\ref{tabSI:openstreetmap}) and two features from charging pool coordinates in the Charging pools 2015 dataset (Table~\ref{tabSI:openchargemap}) by calculating the Euclidean distance to the closest EVnetNL charging pool (first feature) and by calculating the density of points within the buffer areas of EVnetNL charging pools (second feature). Thus, altogether we have created from OpenStreetMap and Charging pools 2015 datasets $32$ features. 

An attribute \textit{Road Segment Type} was associated with each charging pool using the OSMPythonTools library in Python~\cite{Lucas_2018}. To do this, the type of the closest road segment was assigned to a charging pool. The \textit{Road Segment Type} attribute takes one of the following values: residential, primary, secondary and tertiary. We used dummy variables~\cite{James_2014} to derive three features taking zero-one values from the \textit{Road Segment Type} attribute.

In addition, we derived five features from the EVnetNL dataset: \textit{number of charging points}, \textit{maximum power}, \textit{latitude} and \textit{longitude} and the \textit{rollout strategy}. The rollout strategy of the charging pool specifies whether the pool is strategically located or based on the EV user demand. We encode this information with a binary feature. Features derived from the EVnetNL dataset describe technical conditions and the location of the charging pool, and this information may or may not be available in when building a model for a decision making support task, which is why we considered these features as optional. Another reason why we treated these features as optional is given by the fact that they were derived from the same dataset as the response (dependent) variable representing energy consumption. Consequently, there is a higher risk that by including these features we could mask the effect of independent variables derived from external datasets in a regression model. Such a situation is likely to occur if both a response and an independent variable derived from the same datasets are correlated with another independent variable. Such variables are in the literature called confounding variables~\cite[p. 136]{James_2014}. To minimise possible biases, we considered in the analyses two versions of the feature matrix~$\bm{X}$, without and with features derived from the EVnetNL dataset. The differences in the procedure are only minor and to keep the length of the Supplementary information file in a reasonable limit, we further describe the data processing procedure for the feature matrix that includes features derived from the EVnetNL dataset. The feature matrix~$\bm{X}$ with EVnetNL features together with the response vector we attach to the paper.

We identified high level of correlations among attributes in Table~\ref{tabSI:traffic_flows} characterising the average traffic flows as a function of the time of day (i.e. in periods 7-19hr, 19-23hr, 23-7hr). For this reason, we created three features (one for each transport mode) by aggregating the average daily flows: the \textit{number of cars per day} ($TF_{cars} = TF_1 + TF_2 + TF_3$), the \textit{number of buses per day} ($TF_{buses} = TF_4 + TF_5 + TF_6$), and the \textit{number of trucks per day} ($TF_{trucks} = TF_7 + TF_8 + TF_9$). For each charging pool, we identified the closest road segment and used its traffic flow values for each transport mode as a feature. For each buffer area, we calculated the \textit{traffic density} by multiplying the length of all road segments by the traffic flow (considering all three transport modes) and dividing it by the area of the buffer. Besides, we add the \textit{road density} as a feature, \ie the length of the road segments within the buffer area divided by the buffer area~\cite{liu2017evaluation}. Hence, we derived $7$ features from the Traffic Flows dataset.

\begin{figure}
	\centering
	\includegraphics[width=0.7\textwidth]{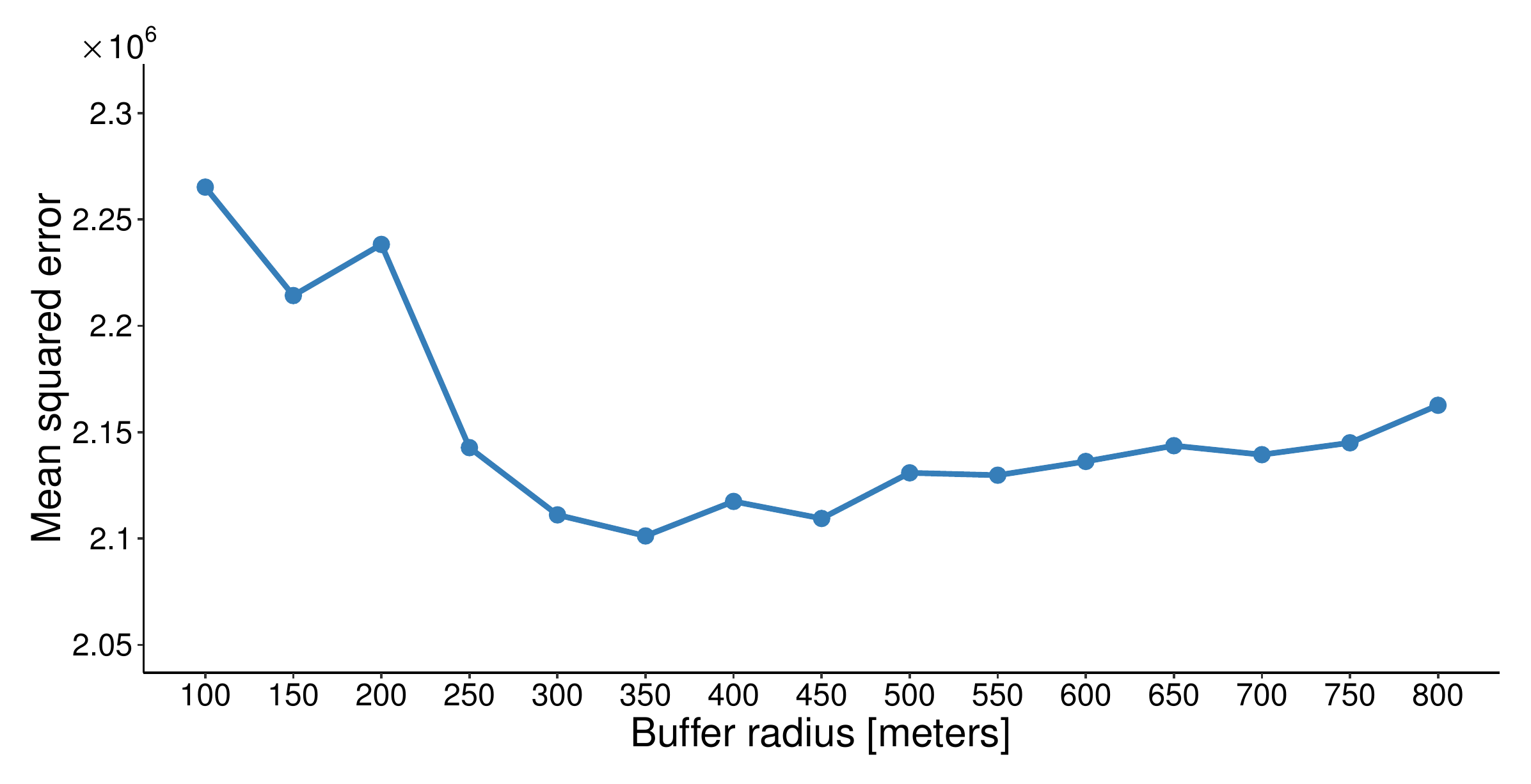}
	\caption{The mean squared error (MSE) of the ordinary least squares (OLS) regression fit of the energy consumed at charging pools during 2015 as a function of the buffer radius. For small radius values, some observations are missing due to the empty intersection of buffers and the Population Cores dataset. To make the results comparable across different values of the radius, we reduced the observations to those obtained for the radius of $100$ meters. The best fit (the smallest MSE) was achieved for the buffer radius of $350$ meters.}
	\label{fig:buffer_radius}
\end{figure}

To determine the buffer radius $r$, we used values ranging from $100$ to $800$ meters incremented by $50$ metres. This range covers the walking distance to reach the final destination after parking the vehicle for a large percentage of trips~\cite{Waerden_2015}. We applied the OLS regression to investigate the fit between the consumed energy on charging pools and GIS features (see Figure~\ref{fig:buffer_radius}). The observed dependence indicates that the buffer radius has a minor impact on the fit, while the best fit we observe for the buffer radius of $r = 350$ meters. For buffers with $r = 350$ meters, we obtained $195$ features.
\subsection{Treatment of potential modelling problems} 
\label{secSI:modelling_problems}
To avoid using uninformative data,  we considered only features with more than $5$\% of non-zero values in the buffer neighbourhood of all charging pools. Consequently, we eliminated $10$ features derived from some land cover categories.

We treated collinearity in two steps. First, we identified groups of features with the absolute value of the Pearson correlation coefficient between each pair of features greater than $0.95$~\cite[p. 47]{Kuhn_2013}. For each group, we kept only one feature as a representative of the group. Representative features together with features that we excluded from the analysis we report in Table~\ref{tab:correlated_features}.
\newpage
\begin{center}
	\begin{longtable}[h]{p{.3\textwidth} p{.6\textwidth}} 
		\toprule
		Representative feature  &  Excluded features\\
		\hline
		\textit{Number of companies, where the gas consumption was measured} ($EC_{9}$). & \textit{Number of companies where was electric energy consumption measured} ($EC_{12}$). \\
		\hline
		\textit{Percentage of owner-occupied properties} ($N_{39}$). &  \textit{Percentage of rental properties} ($N_{40}$).\\
		\hline
		\textit{Percentage of households with low purchasing power} ($N_{51}$). &  \textit{Percentage of households below or around the social minimum, except student households} ($N_{52}$).\\
		\hline
		\textit{Total population} ($N_{4}$). & \textit{Total number of residential properties with at least one residential function and possibly one or more other usage functions} ($N_{35}$); \textit{Number of households} ($N_{16}$); \textit{Number of individuals receiving an income} ($N_{44}$);  \textit{Annual consumption of electric energy of residential properties} ($EC_{5}$); \textit{Number of residential properties where gas consumption was measured} ($EC_{3}$); \textit{Number of residential properties where electric energy was measured} ($EC_{6}$). \\
		\hline
		\textit{Average household size} ($N_{19}$). &  \textit{Percentage of private households with one person} ($N_{17}$).\\
		\hline
		\textit{Average income in thousand EUR per resident} ($N_{46}$.) & \textit{Average income in thousand EUR per an individual receiving income} ($N_{45}$).\\
		\hline
		\textit{Number of individuals aged 0 - 14 in multi-person households without biological or adopted children, or stepchildren} ($PC_{6}$). & \textit{Number of individuals aged 25 - 44 in multi-person households without biological or adopted children, or stepchildren} ($PC_{8}$).\\
		\hline
		\textit{Number of individuals aged 15 - 24 in multi-person households without biological or adopted children, or stepchildren} ($PC_{7}).$ & \textit{Number of individuals aged 45 - 64 in multi-person households without biological or adopted children, or stepchildren} ($PC_{9}$).\\
		\hline
		\textit{Number of individuals living as an unmarried couple or couple without registered civil partnership without children that belong to a household of two people} ($PC_{17}$). & 
		\textit{Number of individuals aged 25 - 44 in multi-person households with biological or adopted children, or stepchildren} ($PC_{13}$). \\ 
		\hline
		\textit{Number of individuals living as a married or couple with registered civil partnership without children that belong to a household of two people} ($PC_{19}$). & \textit{Number of individuals aged 45 - 64 in multi-person households with biological or adopted children, or stepchildren} ($PC_{14}$). \\
		\hline               
		\textit{Number of individuals living in a private household but not in a relationship} ($PC_{21}$). &  \textit{Number of individuals aged 25 - 44 in one-person households} ($PC_{3}$). \\
		\hline
		\textit{Number of households with five individuals} ($PC_{40}$). & \textit{Number of individuals living as a married or couple with registered civil partnership with children} ($PC_{18}$). \\
		\hline
		\textit{Average house value in EUR} ($PC_{42}$). & \textit{Average value of owner-occupied house in EUR} ($PC_{43}$); \textit{Average value of rental housing in EUR} ($PC_{44}$).\\
		\hline
		\hline
		\caption{Each row corresponds to a group of features with mutual values of the Pearson correlation coefficient greater than $0.95$. The feature in the first column was selected as a representative of the group of features listed in the second column.}
		\label{tab:correlated_features}
	\end{longtable}
\end{center}
\newpage
To detect multicollinearity, we evaluated the variance inflation factor (VIF)~\cite{chatterjee2015regression}. We iteratively removed the feature with the largest VIF and recomputed the VIF after each removal, until all VIF values were below 10~\cite{James_2014}. This process led to the elimination of $47$ features, in the following order: $PC_{37}$, $PC_{23}$, $N_{34}$, $PC_6$, $N_{21}$, $N_{63}$, $PC_{38}$, $N_9$, $N_{39}$, $PC_{20}$, $PC_{19}$, $N_4$, $PC_{17}$, $N_8$, $N_{62}$, $PC_{39}$, $PC_4$, $N_{48}$, $N_{49}$, $LC_4$, $PC_{21}$, $PC_{31}$, $PC_5$, $N_{10}$, $PC_{26}$, $N_{32}$, $N_{46}$, $N_6$, $N_2$, $N_{54}$, $PC_7$, $N_{47}$, $L_6$, \textit{traffic density of buses}, $PC_{25}$, $N_{29}$, $N_{33}$, $N_{19}$, $N_{58}$, $PC_{40}$, $PC_{24}$, $EC_2$, $PC_{12}$, $N_{53}$, $N_{50}$, residential road segment, $L_2$. To check also for other measures of multicollinearity, we calculated the eigenvalues of the correlation matrix. The smallest value was equal to $0.04938$ (i.e. greater than the threshold value $0.01$ recommended in the literature) and all eigenvalues sum to $429.11$ ( which is not greater than $5$ times the number of predictors as recommended in the literature)~\cite[p.~252]{chatterjee2015regression}.

\begin{figure}
	\centering
	\includegraphics[width=0.9\textwidth]{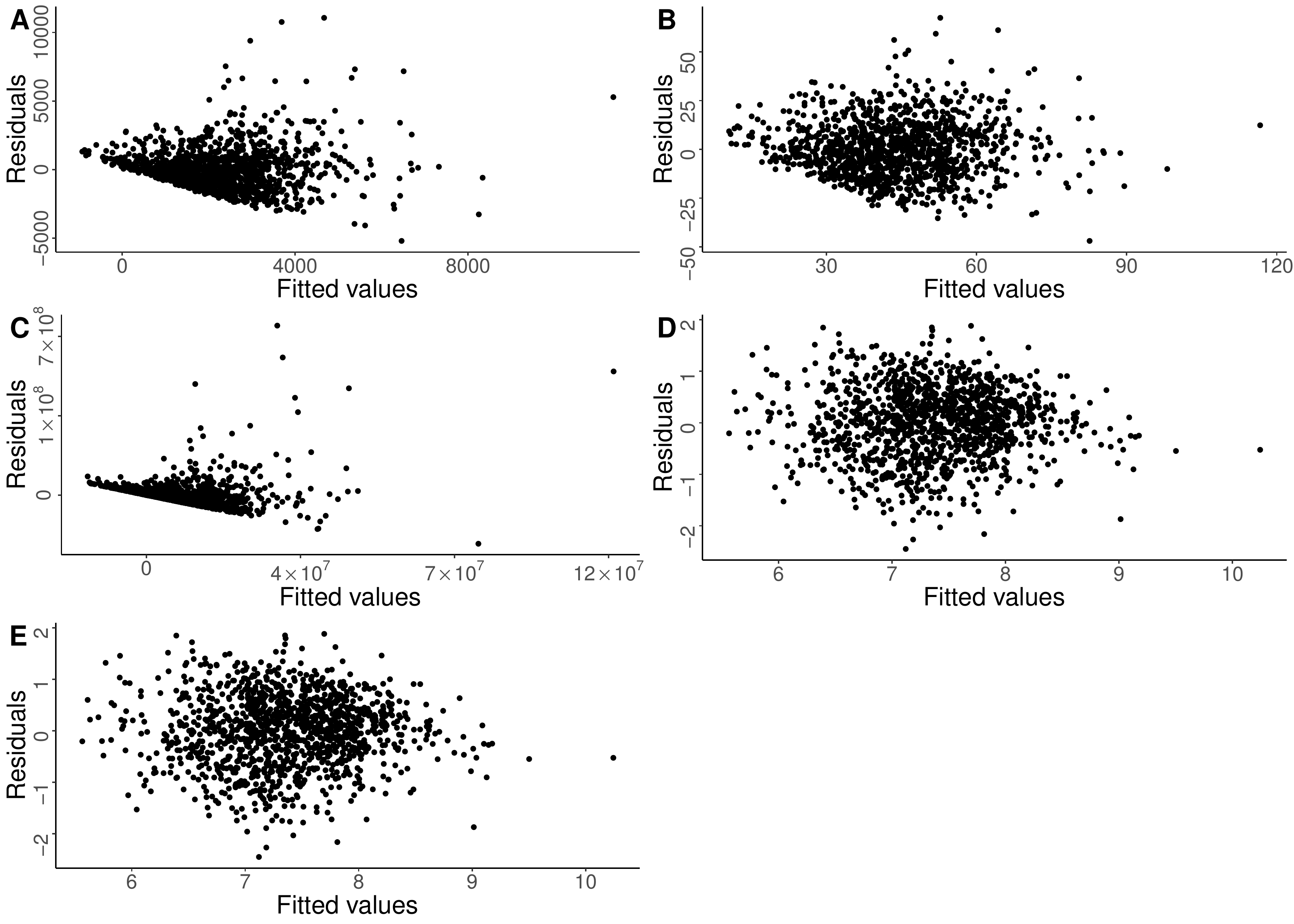}
	\caption{Residual plots. \textbf{A} Residual plot obtained by applying the OLS regression to the feature matrix $\bm{X}$ (after applying the data pre-processing steps) and to the response vector $\bm{y}$, \textbf{B} transformation $\sqrt{\bm{y}}$, \textbf{C} transformation $\bm{y}^2$, \textbf{D} transformation $\log(\bm{y})$ and \textbf{E} Box-Cox transformation (with $\lambda = 0.1$). Residuals appear more random after applying $\log(\bm{y})$ and Box-Cox transformations.}
	\label{fig:residual_plots}
\end{figure}

We tested whether transformations: $\sqrt{\bm{y}}$, $\bm{y}^2$, $\log(\bm{y})$ and the Box-Cox transformation~\cite[p.~32]{Kuhn_2013} of the response vector $\bm{y}$ can improve the linear model obtained by the OLS regression with the feature matrix $\bm{X}$. In Figure~\ref{fig:residual_plots} are shown residual plots. For a suitable model, the residual plot should show no distinguishable pattern, indicating that the model explains the response variable well. On the contrary, the presence of a pattern in the residual plot indicates that introducing some non-linearities could help to improve the model. Residual plots, in panels D and E of Figure~\ref{fig:residual_plots}, indicate that transformation $\log(\bm{y})$ and Box-Cox transformation with the value of the exponent $\lambda = 0.1$, led to the least regularities formed by residuals. The Box-Cox function is in the limit approaching the $\log$ function, when $\lambda$ is going to zero. Hence, a value of $\lambda = 0.1$ that is close to zero, makes both transformations very similar. For this reason, in numerical experiments presented in Section~\ref{sec:energy_consumption_from_urban_data} and~\ref{subsec:rollout_strategy_effects} we applied the transformation $\log(\bm{y})$. 

\begin{figure}
	\centering
	\includegraphics[width=0.9\textwidth]{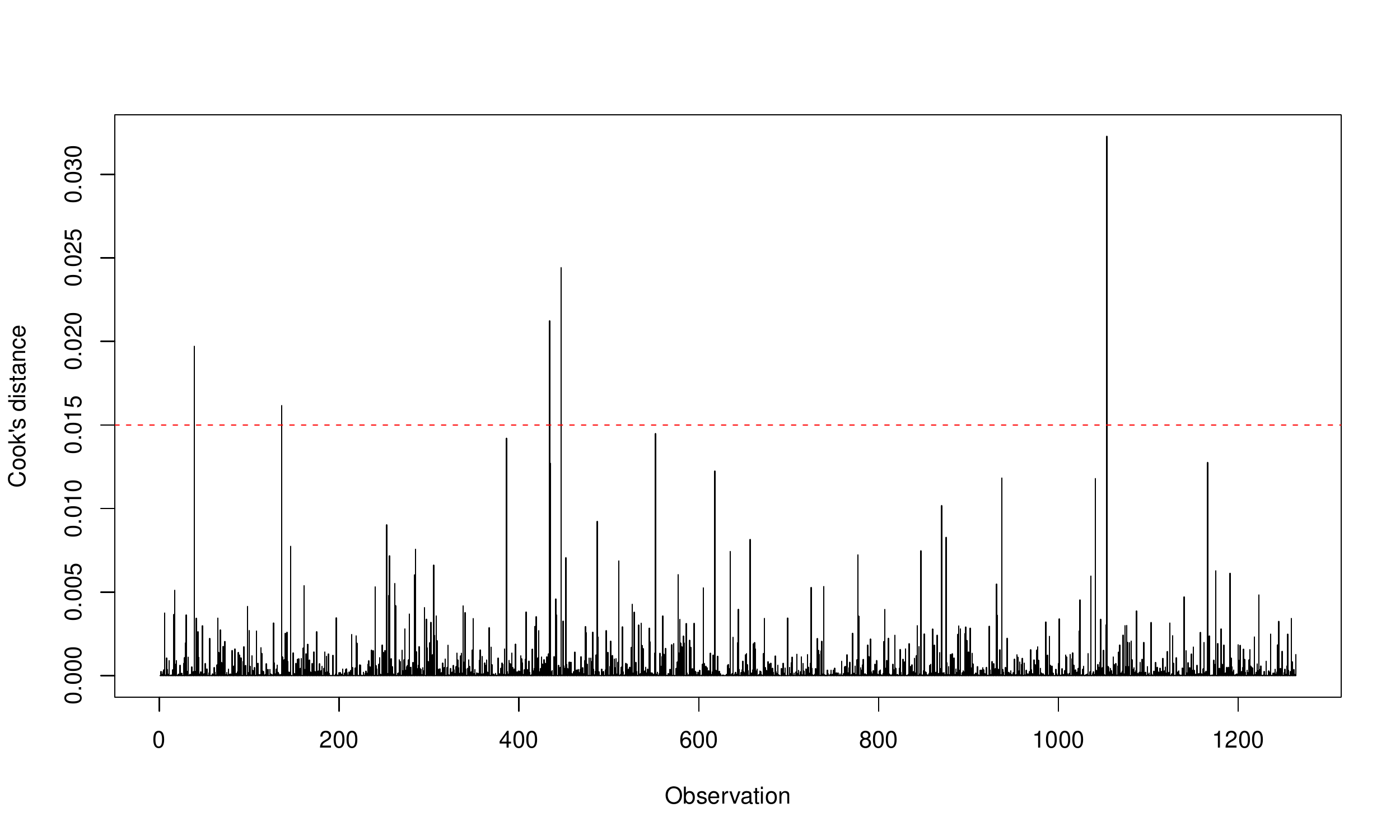}	
	\caption{Visualisation of Cook's distances obtained for individual observations (charging pools). The dashed line indicates the threshold value that was applied to mark influential observations (i.e. those exceeding the threshold value).}
	\label{fig:cooks_plot}
\end{figure}

In the literature, the Cook's distance is recommended~\cite{chatterjee2015regression} to be used to detect observations which cause substantial changes to a regression model when deleted. The Cook's distance can be interpreted as a distance between the quality of the fit when using complete data and the quality of the fit without an observation. In Figure~\ref{fig:cooks_plot}, we calculated values of the Cook's distance for all observations. We estimated values sticking out from the rest (i.e. influential points). As recommended in Ref.~\cite{chatterjee2015regression}, we set the threshold value to $0.015$ and found $5$ observations where the Cook's distance exceeded the threshold. These observations were considered as influential points and excluded from the dataset. The quality of the fit was improved, e.g. the mean square error (MSE) of the OLS regression decreased from $0.523$ to $0.512$, i.e. by $2.4\%$.

Finally, we obtained $1259$ observations and $119$ features derived from geospatial datasets characterizing the vicinity of charging pools and $5$ features derived from EVnetNL dataset specifying the location and basic characteristics of charging pools. 

\section{Fitting the energy consumption with a transformed probability density function}
\label{secSI:energyFitting}
\begin{table}[]
		\begin{tabular}{p{0.05\textwidth}p{0.15\textwidth}p{0.15\textwidth}p{0.15\textwidth}p{0.15\textwidth}}
			&  & \multicolumn{3}{c}{Distribution} \\ \cline{3-5} 
			&  & Weibull     & beta    & gamma    \\ \cline{2-5} 
			\multirow{6}{*}{\rotatebox{90}{Transf., $g(\bm{y})$}} & $\bm{y}$ & 0.093 & 0.000 & 0.026  \\ 
			&  $\bm{y}^2$         & 0.013 & 0.000 & 0.000       \\ 
			&  $\bm{y}^3$         & 0.019 & 0.000 & 0.000 \\ 
			&  $\sqrt{\bm{y}}$    & 0.093 & 0.924 & 0.163 \\ 
			&  $\sqrt[3]{\bm{y}}$ & 0.093 & 0.582 & 0.095 \\ 
			&  $log(\bm{y})$      & 0.506 & 0.081 & 0.000 \\ \cline{2-5}
			
		\end{tabular}
		
		\caption{Results of the Kolmogorov-Smirnov goodness-of-fit test (p-values) obtained after applying the test to the response vector $\bm{y}$ and the probability density function $f_Y(y)$. Considering the transformation $Z = g(Y)$, the density $f_Y(y)$ is obtained from $f_Z(z)$. The density $f_Z(z)$ was fitted to energy consumption data using one of the standard distributions (Weibull, beta and gamma). In several cases, we found statistically significant results (p-value $> 0.05$). We ran tests also with exponential, normal and log-normal probability density functions finding only non-significant results.}
	\label{tabSI:KS_test}
\end{table}
Let $Y$ be the random variable, with the probability density function $f_Y(y)$, modelling the distribution of the energy consumption at charging pools. To estimate the density $f_Y(y)$, we considered a transformation $g(\bm{y})$ of the response vector $\bm{y}$ (the function $g()$ was applied to $\bm{y}$ elementwise) and we defined random variable $Z = g(Y)$ described by the probability density function $f_Z(z)$. Hence, the random variable $Z$ is modelling the vector $\bm{z} = g(\bm{y})$. To fit the vector $\bm{z}$ with  $f_Z(z)$, we considered three standard probability distributions (beta, gamma, and Weibull). Note that for the beta distribution, the response vector $\bm{y}$ needs to be re-scaled to the interval $\langle0, 1\rangle$. We used the maximum likelihood~\cite[p.~122]{wasserman2013all} and the method of moments~\cite[p.~120]{wasserman2013all} for fitting. The method of moments gave a better fit for beta and gamma distributions, while for the Weibull distribution, we obtained better fits with the maximum likelihood method. 
\begin{figure}[]
	\centering
	\includegraphics[width=0.9\textwidth]{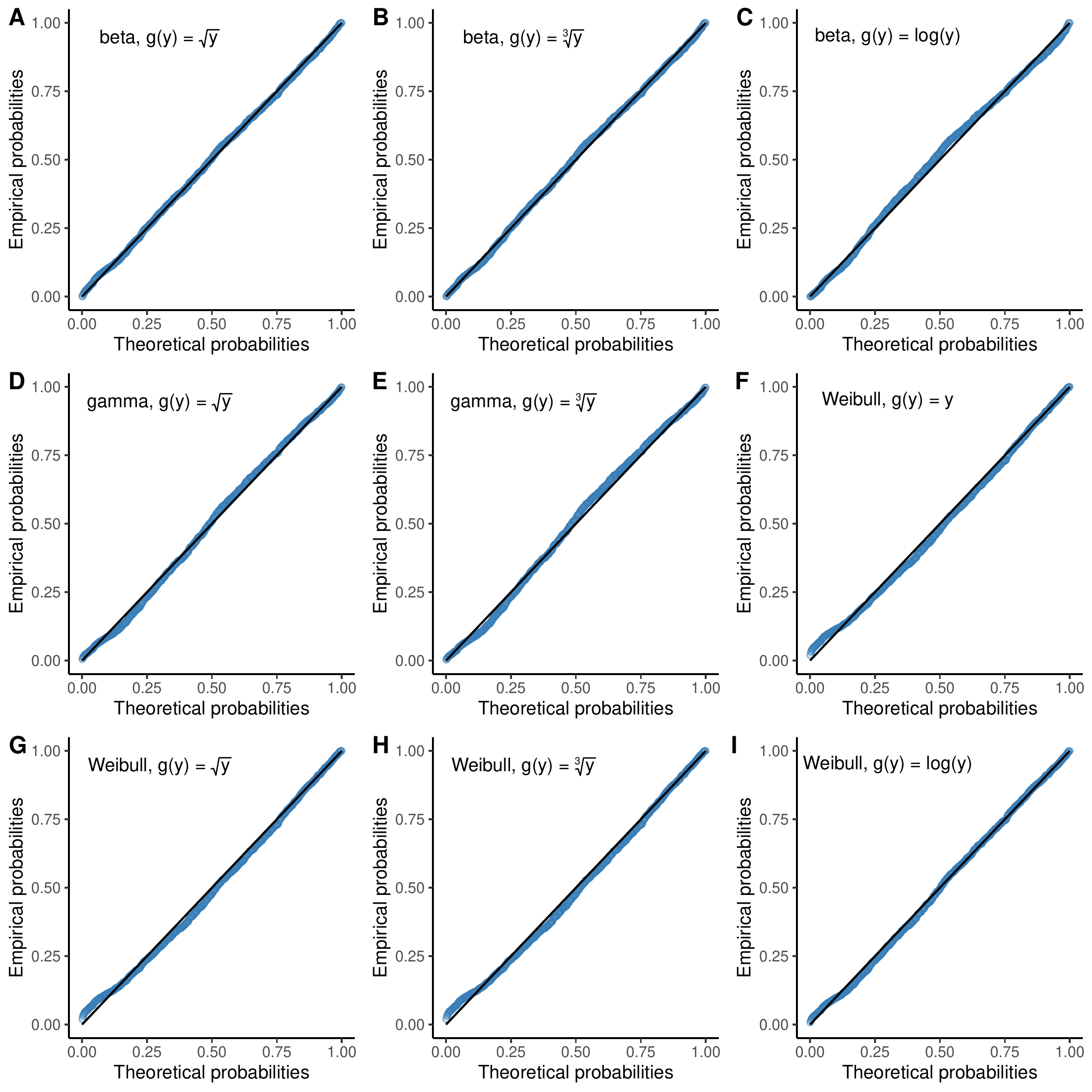}	
	\caption{The P-P plots for statistically significant (p-value $> 0.05$) combinations of the distribution function with the transformation $g(\bm{y})$ from Table~\ref{tabSI:KS_test}. In the upper left corner of each panel is indicated the used distribution function and the transformation function $g(\bm{y})$. The closer the points are to the diagonal line, the better the probability density function $f_Y(y)$ approximates the response vector $\bm{y}$.}
	\label{figSI:pp_plots}
\end{figure}

\begin{figure}[]
	\centering
	\includegraphics[width=0.9\textwidth]{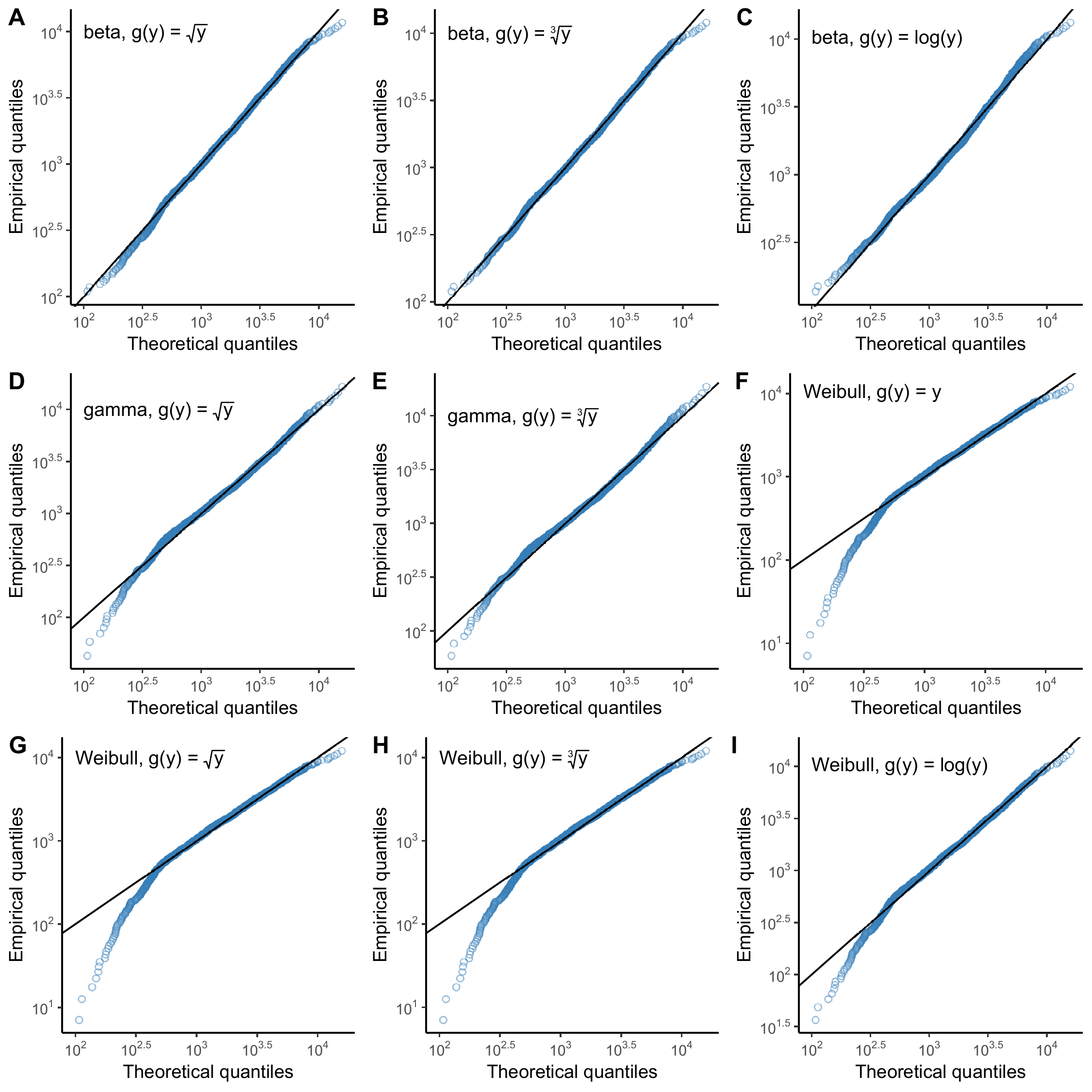}	
	\caption{The Q-Q plots (on log-log scale) for statistically significant (p-value $> 0.05$) combinations of the distribution function with the transformation $g(\bm{y})$ from Table~\ref{tabSI:KS_test}. In the upper left corner of each panel is indicated the used distribution function and the transformation function $g(\bm{y})$. The closer the points are to the diagonal line, the better the probability density function $f_Y(y)$  approximates the response vector $\bm{y}$.}
	\label{figSI:qq_plots}
\end{figure}

In Table~\ref{tabSI:KS_test}, we report the p-values obtained by the Kolmogorov-Smirnov test applied to $f_Y(y)$ and $\bm{y}$, while considering several transformations $g(\bm{y})$. To explore which values of the response vector $\bm{y}$ are better explained with a probability density function $f_Y(y)$, it is recommended to combine the Kolmogorov-Smirnov goodness-of-fit test with a graphical method~\cite{warren2010application}, e.g. the P-P or the Q-Q plot. In Figures~\ref{figSI:pp_plots} and~\ref{figSI:qq_plots}, we show the P-P and the Q-Q plots obtained from $\bm{y}$ and $f_Y(y)$ for selected distribution functions and for transformations that resulted in Table~\ref{tabSI:KS_test} in a statistically significant test. 

There is no single distribution capturing the data in an ideal way in the whole range of values. The largest p-values we found for $g(\bm{y}) = \sqrt{\bm{y}}$ and $g(\bm{y}) = \sqrt[3]{\bm{y}}$ combined with the beta distribution and $g(\bm{y}) =\log(\bm{y})$ combined with the Weibull distribution which are shown in panels A, B and I of Figures~\ref{figSI:pp_plots} and~\ref{figSI:qq_plots}. The Q-Q plots indicate that the beta distribution explains better small values than the Weibull distribution. A few largest values are not captured well, neither by the beta nor by the Weibull distribution. Although the $g(\bm{y}) = \sqrt{\bm{y}}$ combined with beta distribution has larger p-value than $g(\bm{y}) = \sqrt[3]{\bm{y}}$, the Q-Q plot indicates that it fits worse the charging pools with low energy consumption. Thus, considering p-values, the P-P plots and the Q-Q plots, we concluded as the most satisfactory fit the beta distribution (with parameter values $\alpha = 2.576$ and $\beta = 4.528$) combined with the transformation $g(\bm{y}) = \sqrt[3]{\bm{y}}$. 
\section{Extended results}
\label{secSI:inferenceResults}
\begin{figure}[]
	\centering
	\includegraphics[width=0.9\textwidth]{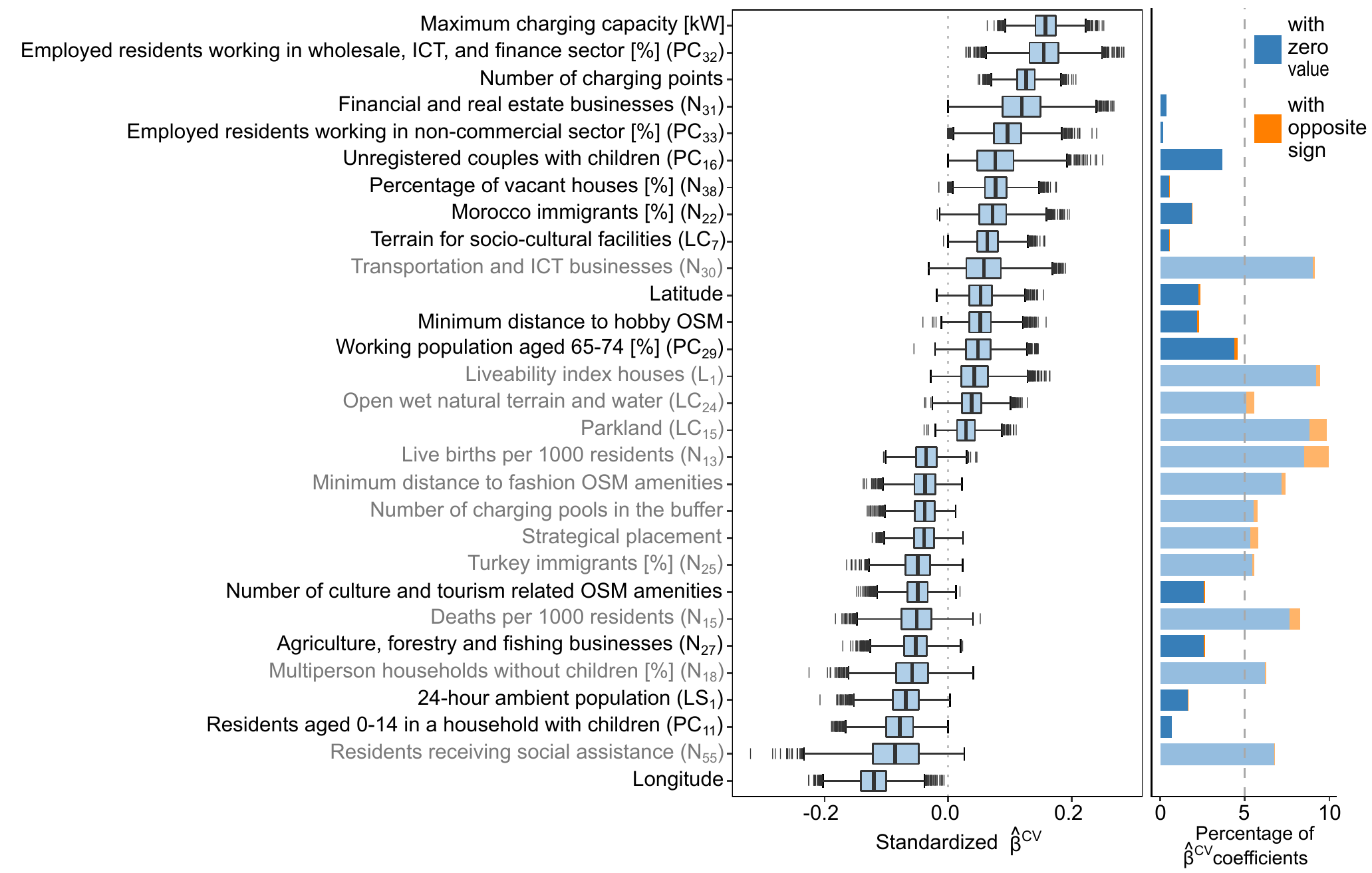}	
	\caption{The empirical distributions of standardized regression coefficients obtained by the Lasso method and the $10$-fold cross-validation applied to $10~000$ samples of bootstrapped data. To explain consumed energy, we included features derived from the EVnetNL dataset. We show only features where the value of the regression coefficient was set to zero in less than $10\%$ of the samples. Coefficients are descendingly ordered, from the largest to the smallest median value. The left panel presents the Tukey's box plot of coefficients. On the right, the stacked bar plot shows the percentage of samples when the regression coefficient $\hat{\beta}^{CV}$ was set to zero and the number of samples where it reached the opposite sign as the sign of the median. We consider as significant those features where the number of samples with zero coefficient value is less than $5\%$ and the number of samples with opposite sign is small. The dashed line indicates the $5\%$ threshold value. Features that are not considered significant we display using faded colours. Full descriptions of coefficients can be found in tables~\ref{tabSI:Population_cores_attr}~-~\ref{tabSI:landscan} of the SI file by using the code in brackets.}
	\label{figSI:energPoolFeatures}
\end{figure}

\begin{figure}[]
	\centering
	\includegraphics[width=0.8\textwidth]{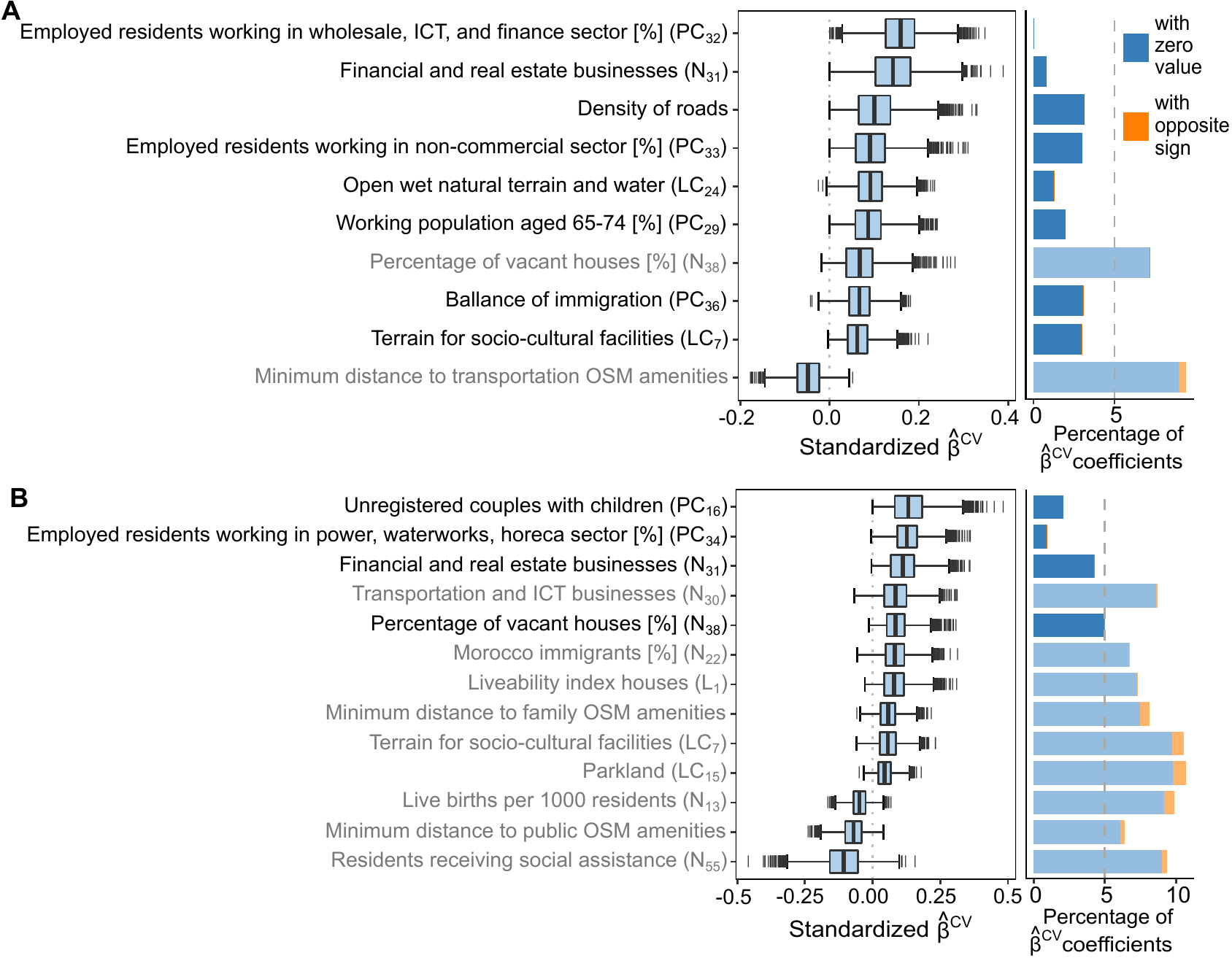}	
	\caption{
		The empirical distributions of standardized regression coefficients of two models (each built for a different group of charging pools). Each charging pool is assigned to one group depended on whether it is located in a municipality populated with less (panel \textbf{A}) or more (panel \textbf{B}) than $50~000$ residents. The results were obtained by the Lasso method combined with the $10$-fold cross-validation applied to $10~000$ samples of bootstrapped data. 
		We display only features where the value of the regression coefficient was set to zero in less than $10\%$ of the samples. Regression coefficients are descendingly ordered, from the largest to the smallest median value. The left panel presents the Tukey's box plot of coefficients. On the right, the stacked bar plot shows the percentage of samples when the regression coefficient $\hat{\beta}^{CV}$ was set to zero and the number of samples where it reached the opposite sign as the sign of the median. We consider those features as significant where the number of samples with zero coefficient value is less than $5\%$ and the number of samples with opposite sign is small. The dashed line indicates the $5\%$ threshold value. Features that are not considered significant we display using faded colours. Full descriptions of coefficients can be found in tables~\ref{tabSI:Population_cores_attr}~-~\ref{tabSI:landscan} of the SI file by using the code in brackets.}
	\label{figSI:energyCities}
\end{figure}

Figure~\ref{figSI:energPoolFeatures} displays the empirical distributions of standardized regression coefficients of the model that is fitting energy consumption with geospatial features together with features derived from the EVnetNL dataset. Comparison of Figure~\ref{fig:coef_imp_lasso_boot} with Figure~\ref{figSI:energPoolFeatures} reveals that some of the features in the model were replaced by the features derived from the EVnetNL dataset, pointing out the relevance of the charging pool characteristics for energy consumption. Figure~\ref{figSI:energyCities} displays the distributions of standardized regression coefficients of two models that are fitting energy consumption at two groups of charging pools. Each charging pools is assigned to one group depended on whether it is located in a municipality populated with less (Figure~\ref{figSI:energyCities}A) or more (Figure~\ref{figSI:energyCities}B) than $50~000$ residents.

\end{document}